\documentclass[aps,prb,twocolumn,floatfix,footinbib,showpacs,superscriptaddress]{revtex4}
\usepackage{graphicx}
\usepackage{amsfonts,amsmath,amssymb}
\usepackage{amsthm}
\usepackage{dsfont,bm}
\usepackage{color}
\usepackage{soul} 
\usepackage{amsbsy}
\usepackage[colorlinks=true,linkcolor=blue,pagecolor=blue,filecolor=blue,menucolor=blue,urlcolor=blue,citecolor=blue,anchorcolor=blue]{hyperref}%
\usepackage{sidecap}
\usepackage{txfonts}
\usepackage{amsbsy}
\usepackage{times} 
\usepackage{dcolumn}

\newcommand{\tw}{2$\times$2$\times$2}
\newcommand{\tr}{3$\times$3$\times$3}
\newcommand{\fr}{4$\times$4$\times$4}

\begin{document}

\title{Density functional simulations of pressurized Mg-Zn and Al-Zn alloys}
\author{Mohammad Alidoust}
\affiliation{Department of Physics, Norwegian University of Science and Technology (NTNU), NO-7491, Trondheim, Norway}
\author{David Kleiven}
\affiliation{Department of Physics, Norwegian University of Science and Technology (NTNU), NO-7491, Trondheim, Norway}
\author{Jaakko Akola}
\affiliation{Department of Physics, Norwegian University of Science and Technology (NTNU), NO-7491, Trondheim, Norway}
\affiliation{Computational Physics Laboratory, Tampere University, FI-30014, Tampere, Finland}

\date{\today} 
\begin{abstract} 
The Mg-Zn and Al-Zn binary alloys have been investigated theoretically under static isotropic pressure. The stable phases of these binaries on both initially hexagonal-close-packed (HCP) and face-centered-cubic (FCC) lattices have been determined by utilizing an iterative approach that uses a configurational cluster expansion method, Monte Carlo search algorithm, and density functional theory (DFT) calculations. Based on 64-atom models, it is shown that the most stable phases of the Mg-Zn binary alloy under ambient condition are $\rm MgZn_3$, $\rm Mg_{19}Zn_{45}$, $\rm MgZn$, and $\rm Mg_{34}Zn_{30}$ for the HCP, and $\rm MgZn_3$ and $\rm MgZn$ for the FCC lattice, whereas the Al-Zn binary is energetically unfavorable throughout the entire composition range for both the HCP and FCC lattices under all conditions. By applying an isotropic pressure in the HCP lattice, $\rm Mg_{19}Zn_{45}$ turns into an unstable phase at P$\approx$$10$~GPa,  a new stable phase $\rm Mg_{3}Zn$ appears at P$\gtrsim$$20$~GPa, and $\rm Mg_{34}Zn_{30}$ becomes unstable for P$\gtrsim$$30$~GPa. For FCC lattice, the $\rm Mg_{3}Zn$ phase weakly touches the convex hull  at P$\gtrsim$$20$~GPa while the other stable phases remain intact up to $\approx$$120$~GPa. Furthermore, making use of the obtained DFT results, bulk modulus has been computed for several compositions up to pressure values of the order of $\approx$$120$~GPa. The findings suggest that one can switch between $\rm Mg$-rich and $\rm Zn$-rich early-stage clusters simply by applying external pressure. $\rm Zn$-rich alloys and precipitates are more favorable in terms of stiffness and stability against external deformation.    
\end{abstract}
\maketitle

\section{introduction} \label{introduction}

The controlled creation of alloys is an attractive research field from both theoretical and experimental standpoints. \cite{J.F.Nie,T.M.Pollock,F.W.Gayle} An alloy system comprises numerous charged particles that interact with each other and make a highly complicated interacting many-body platform. As it is evident, this level of complexity makes accurate theoretical studies of such systems quite challenging, and, therefore multiple approximations with today's computers have to be incorporated. \cite{D.E.Dickel,davidkl,Z.Wu,Y.-M.Kim,P.Mao,B.Zheng,T.Tsurua,C.Ravi,A.V.Ruban,A.Yu.Stroev,B.Zou} The interplay of the quantum mechanical interactions and the specific spatial ordering of different atoms, forming an alloy, determines the macroscopic properties of alloys, and this has served as a unique platform for testing and developing various theoretical approaches. \cite{D.E.Dickel,Z.Wu,Y.-M.Kim,J.M.Sanchez,H.S.Jang,M.I.Baskes} On the other hand, the detailed atomistic level insight from theoretical studies may provide useful information for synthesizing alloys in experiments with boosted specific properties. \cite{S.Wang,Y.Z.Ji,T.E.M.Staab,S.R.Nayak,A.Yu.Stroev} Prominent examples can be found in aluminium alloys that play increasingly crucial roles in various industrial products. \cite{C.Wang,davidkl,B.Zheng,Y.-M.Kim,T.Tsurua,C.Ravi,G.Yi,M.Liu,G.Esteban-Manzanares,B.Zou}

One of the main goals of alloying matrices is to obtain the most lightweight and high-strength alloys that can be extensively used in aircraft industries, modern trains and vehicles, and so forth. The central mechanisms for enhancing the strength of aluminium is to produce uniform distributions of nucleated precipitates and solid microstructures in the aluminium matrix with different solute elements such as $\rm Zn$, $\rm Mg$, $\rm Si$, $\rm Cu$, $\rm Mn$, and $\rm Cr$. \cite{S.J.Andersen,T.Saito,S.Wenner,W.Chrominski,E.Christiansen,Y.Xin,X.Gao,W.Lefebvre,J.F.Nie2,H.D.Zhao} Among them, however, $\rm Zn$ and $\rm Mg$ have demonstrated to be the most effective additives to help the growth of precipitates from solute clusters, thus enhancing the strength and hardness of the $\rm Al$ matrix through generating high resistance against dislocation motion and damage. For example, a recent experiment has demonstrated that a cyclic deformation of an $\rm Al$ alloy at room temperature facilitates a diffusion of solid solute clusters and eventually generates a matrix with uniformly close packed precipitates that increase the material strength and elongation properties. The advantages of this method compared to conventional temperature-aging approaches are (i) a shorter processing time and (ii) a more uniformly distributed microstructure with no precipitate-free zones \cite{W.Sun}. Also, it has been experimentally observed that several phases of precipitates can develop in Mg-Zn and Al-Zn-Mg alloys, including $\rm MgZn$, monoclinic $\rm Mg_4 Zn_7$, $\eta$-$\rm MgZn_2$, $\beta$-$\rm MgZn_2$ with hexagonal symmetry, $\rm Mg_2Zn_3$, and $\rm Mg_2Zn_{11}$ \cite{A.Bendo,A.Singh,A.Lervik,T.F.Chung,W.Liu,H.Okamoto,J.Peng}. Recent systematic investigations have demonstrated that the morphology and structural evolution of precipitates in $\rm Mg$ alloys are closely linked to their internal sub-unit cell arrangements, the aspect ratio of precipitates, misfit strains, and their interaction with matrix interfaces \cite{A.Bendo,E.L.S.Solomon,S.DeWitt,B.Sun}. 

Despite the extensive attention that aluminium alloys have received so far, only limited studies have been performed for investigating their fundamental properties when subject to tension from atomistic level first-principles. \cite{J.Y.Wang,P.Ma} The findings of such high-end rigorous studies not only can be used as guidelines for future experiments but they also can be employed as benchmark tests for computationally less expensive approaches (such as effective potential methods \cite{D.E.Dickel,Z.Wu,Y.-M.Kim,M.I.Baskes,H.S.Jang} which are crucial for molecular dynamics simulations or phase field approximations). Hence, these results can serve in developing reliable multiscale models for providing better insights into closer-to-realistic sample sizes. Also, there are several magnesium and zinc rich alloys that, respectively, contain $\rm Al$ / $\rm Zn$ and $\rm Al$ / $\rm Mg$ elements as the most effective additions to improve their hardness. Therefore, it is of fundamental interest to study these alloys within the entire composition range. In order to provide insights for the Al-Mg-Zn alloys, one first needs to study the properties of the Al-Mg, Mg-Zn, and Al-Zn binaries, which comprise the boundaries of the Al-Mg-Zn trinary composition diagram. The properties of the Al-Mg binary have already been studied in previous works \cite{davidkl,B.Zou}, and thus, we focus here on the Mg-Zn and Al-Zn binaries only. 

In this work, we employ an iterative combination of a cluster expansion (CE) method, Monte Carlo (MC) simulations, and DFT calculations to exhaustively search and find the stable phases of the Mg-Zn and Al-Zn binary alloys, starting from both hexagonal-close-packed (HCP) and face-centered-cubic (FCC) lattice symmetries. Our theoretical results show that the stable phases of the Mg-Zn binary are mixture and layered configurations of the two elements in the HCP lattice, and they are in line with experimental observations where $\rm MgZn_2$, $\rm Mg_4Zn_7$, and $\rm MgZn$ compositions with hexagonal symmetry have been found as the most stable phases. For the FCC lattice, we find layered geometrical motives emerging for $\rm MgZn$ and higher Mg-concentrations. The Al-Zn binary turns out to be energetically unfavorable on both the HCP and FCC lattice symmetries under all pressure conditions studied, which indicates a segregation behavior. Although the free energies of $\rm Mg$-rich compositions are lower than those of $\rm Zn$-rich ones for the Mg-Zn binary, the presence of an external pressure can revert the situation such that $\rm Mg$-rich alloys become unstable. The analysis of bulk modulus for several compositions illustrates that $\rm Zn$-rich compositions have the largest bulk modulus and they are more resistant to external deformation. These findings suggest that by applying an appropriate pressure to early-stage clusters of the Mg-Zn binary, one can control the type of the precipitate clusters (being either $\rm Zn$-rich or $\rm Mg$-rich solid solutions), and thus, tune the absolute hardness of the entire matrix.

The article is organized as follows. In Sec. \ref{method}, we briefly discuss and present the theoretical framework used in order to find the energetically favorable phases of binary alloys. In Sec. \ref{results}, we present our main findings and a detailed description of the results. Finally, in Sec. \ref{conclusions}, we present concluding remarks.

\section{Methods}\label{method}

In this section, we describe the theoretical framework and computational details. 

\subsection{ Density functional theory calculations}\label{dft}
The first-principles calculations based on the density functional theory of electronic structure were performed using the $\rm GPAW$ code \cite{gp1,gp2}. The interaction between ions (nuclei + core electrons) and valence electrons was described by using the projector augmented-wave method. The gradient-corrected functional by Perdew-Burke-Ernzerhof (PBE) was employed for the exchange correlation energy, the plane-wave cut-off for kinetic energy was set to $600$~eV, and the number of electronic bands was chosen as $120\%$ (corresponding to 20\% unoccupied). Furthermore, we have used 3.5 ${\bf k}$-points per $\AA^{-1}$ in order to grid the ${\bf k}$-space on the basis of the Monkhorst-Pack scheme. In order to produce consistent data, we have kept fixed all the above-mentioned parameters throughout all DFT calculations. The parameter values were chosen based on extensive convergence and consistency tests, and they provide a balance between convergence and good accuracy in all calculations. We have exploited the `fast inertial relaxation engine' and $\rm ExpCellFilter$ modules available in the atomistic simulation environment package \cite{ase} to relax the atomic positions in a given structure under isotropic pressure through self-consistent loops until the residual forces on each atom become less than $0.07$~eV/$\AA$. We have cross-checked our DFT calculations for selected compositions with tighter convergence criteria and found no changes in the resulting convex hull. In total, 861 individual configurations of 64-atoms were DFT-optimized based on these specifications at zero pressure, and the most stable configurations at the convex hull were subjected to additional pressure calculations.

\subsection{Cluster expansion model}\label{ce}
The cluster expansion (CE) model is a generic approach that can describe multi-component systems through trainable parameters. The main objective is thus to develop CE models with the ability to predict configurational properties with high-accuracy and much less computational cost than for DFT. \cite{J.M.Sanchez,D.B.Laks,M.Asta,J.H.Chang,J.Zhang,O.I.Gorbatov,M.Yu.Lavrentiev,J.S.Wrobel} In this work, CE provides an effective Hamiltonian, describing the configurational energy of Mg-Zn and Al-Zn alloys. Each element atom type is represented by a pseudospin $\tau_\alpha$ with an effective cluster interaction $E_\alpha$ at a configurational ordering $\alpha$ (specific order of atoms in a given composition). Thus, with this approach, we can expand and express the energy of a given configuration by: 
\begin{equation}
E = E_0 + \sum_i E_i \tau_i + \sum_{i,j} E_{ij}\tau_i\tau_j + \sum_{i,j,k} E_{ijk} \tau_i\tau_j \tau_k + \dddot ~~.
\end{equation}
The effective cluster interactions $E_\alpha$ are unknown coefficients. The main task is to train the above expansion to DFT data  (training set) and to find the best values for $E_\alpha$ in a truncated series to describe the configurational energy accurately. The accuracy of CE predictions depends crucially on several ingredients: The number of structures in the data set, how the structures are chosen, the number of effective correlation functions (how many terms the expansion possesses), and the correlation functions used for fitting to the data set. The method we chose to find the best truncated series is the minimization of the leave-one-out cross validation (CV) score through the least absolute shrinkage and selection operator (LASSO), i.e.,
\begin{equation}\label{lasso}
\underset{ \alpha}{\min} \Bigg\{  \sum_{n=1}^{N} \left( E_\text{DFT}^n - E_\text{CE}^n \right)^2 + \alpha \sum_\alpha \big| E_\alpha \big|  \Bigg\},
\end{equation}
in which $N$ is equal to the number of configurations evaluated by DFT calculations, $E_\text{DFT}^n$ represents the energy of $n$th structure predicted by DFT, and $E_\text{CE}^n$ is the prediction made by CE. The second term in Eq. (\ref{lasso}) is a cost function, where term $\alpha$ regularizes the coefficients such that it imposes penalty to coefficients with large values and helps in reducing over-fitting. To construct CE models, we have employed the open source PYTHON modules in the CLEASE software. \cite{J.H.Chang}

\begin{figure*}[t]
\begin{tabular}{cc}
\includegraphics[width=7.2cm,height=6.80cm]{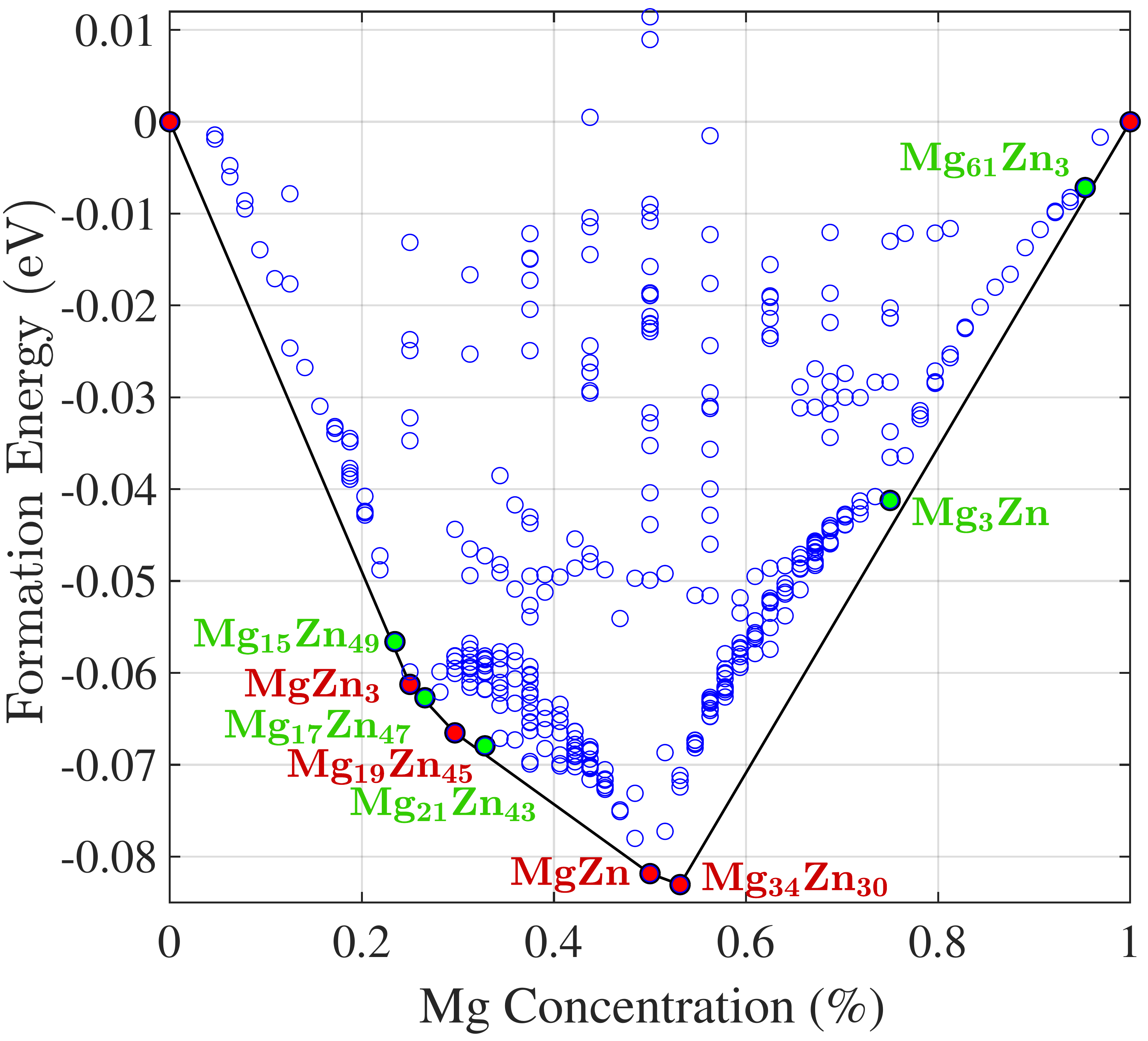} &
\includegraphics[width=10.5cm,height=7.05cm]{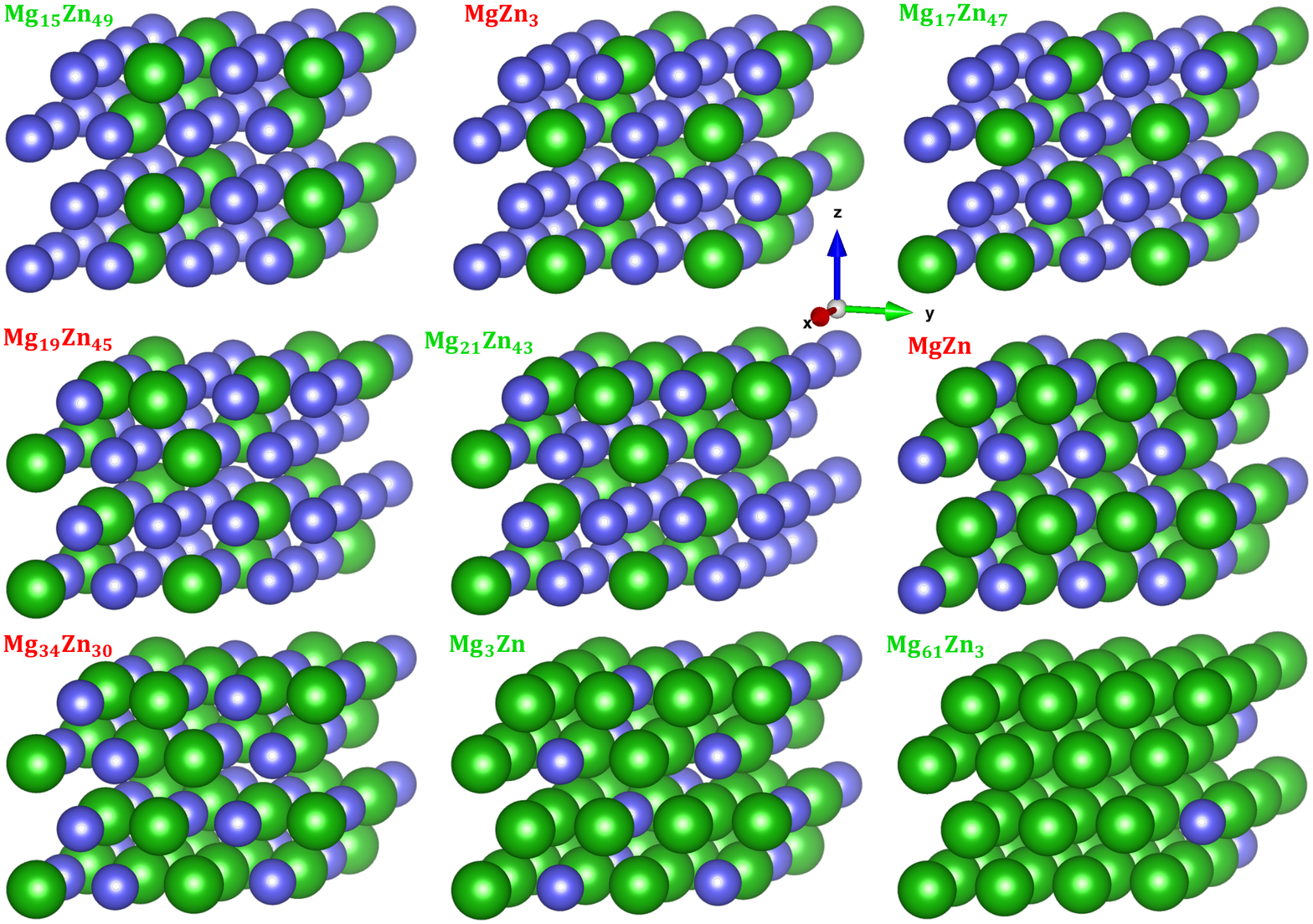} 
\end{tabular}
\caption{ (Color online). Left: The convex hull of the Mg-Zn binary in the HCP lattice symmetry. To obtain the convex hull, we have performed DFT calculations for 402 individual configurations predicted by an iterative use of DFT, CE method and Monte Carlo simulations. The convex hull illustrates that $\rm MgZn_{3}$, $\rm Mg_{19}Zn_{45}$, $\rm MgZn$, and $\rm Mg_{34}Zn_{30}$ (marked by the solid red circles) are stable phases. We have also marked other compositions that are close to the convex hull by the solid green circles. Right: $64$-atom HCP configurations ($4$$\times $$4$$\times $$4$ structures) corresponding to the phases marked by the solid circles in the left panel. The green and blue spheres stand for $\rm Mg$ and $\rm Zn$ atoms, respectively. 
 }
 \label{fig1}
\end{figure*}
 
\subsection{Monte Carlo simulations}\label{mc} 
The configuration space of 64-atom model structures for Mg-Zn and Al-Zn alloys contains a large number of unique configurations, made of various combinations of the considered elements within the entire composition range. Therefore, it is not feasible to evaluate all possible configurations in the configuration space using DFT calculations. In order to effectively search the configuration space and to find the energetically most favorable configurations, we have used the Metropolis algorithm and Monte Carlo simulations \cite{N.Metropolis}. In short, one shuffles the list of atoms in a given configuration and calculates the energy difference between the new $j$ and old $i$ configuration: $\Delta E = E_j - E_i$. Next, one picks up a random number $r$ and defines a probability for accepting the new configuration $j$ together with a simulated annealing algorithm at a temperature $T$ as follows:
\begin{equation}
r  <  \exp \Bigg\{ -\gamma \frac{\Delta E}{k_\text{B} T} \Bigg\}.
\end{equation}  
Here, $k_\text{B}$ is the Boltzmann constant and $\gamma$ is a parameter that controls the acceptance rate of new structures. Depending on the number of possibilities available within the configuration space, one produces a sufficient number of random structures by swapping atoms in each Metropolis step and then increases $\gamma$ slowly from highly small values (equivalent to a high acceptance rate) to large values of $\gamma$ where the acceptance rate of new structures become highly suppressed \cite{N.Metropolis}. 
Note that, at a low enough temperature regime where $k_\text{B}T \rightarrow 0$, one can remove $\gamma$ and use $k_\text{B}T$ as the controlling parameter for the acceptance rate by varying the temperature exponentially from high (equivalent to small values for $\gamma$) to low enough values (equivalent to large values for $\gamma$).

\section{results and discussions} \label{results}
\subsection{Atomic structure, energetics and the effect of pressure}\label{dos}
A CE model is inevitably tied to the elements in question and lattice symmetry, making it almost impractical to train only one model for a set of alloys with different lattice symmetries. Therefore, we have trained several CE models for each alloy system with a specific lattice structure. To find the energetically most favorable structures for Mg-Zn and Al-Zn binaries, we have chosen the following strategy which is tedious but reliable. (i) We produce a database of 40 randomly generated unique structures. (ii) Considering the parameters described in Sec. \ref{dft}, we perform DFT calculations (always requiring geometry and unit cell optimization) for structures generated in the previous stage. (iii) We construct more than 50 CE models as explained in Sec. \ref{ce} by varying the maximum diameter of clusters. (iv) By selecting CE models possessing CV scores (prediction error with respect to DFT results) less than $10$~meV, we generate new unique low-energy structures and add them to the database if not already existing. (v) Extracting these new unique structures, we perform DFT calculations and repeat stages (iii) to (v) until new CE models are not able to add new unique structures into the database, and the procedure reaches convergence.

We have carried out the above steps for the Mg-Zn and Al-Zn binaries on both the HCP and FCC lattice symmetries and used identical fitting procedures for the surrogate CE models. In the following, we have neglected vibrational effects at finite temperature. For the \tw, \tr, and \fr~unit cell structures the concentration of elements changes by a step of $12.50\%$, $3.7037\%$, and $1.5625\%$, respectively. Therefore, the last case (64 atoms) with the smallest steps allows for exploring the full range of concentrations more precisely. We remark here that the structures have {\it initially} an ideal HCP/FCC lattice symmetry but the DFT-optimization (geometry, cell) results in small deviations both in terms of symmetry and individual atomic positions. This means that the DFT total energies present in our CE training database correspond to fully optimized configurations.

In order to compare different compositions based on the (free) energies calculated by DFT, we define the formation energy of a binary ${\cal A}{\cal B}$, regardless of its lattice symmetry, as follows;

\begin{equation}\label{eq1}
H[{\cal A}_x{\cal B}_{1-x}] =  E[{\cal A}_x{\cal B}_{1-x}] - (1-x)E[{\cal B} ]- xE[{\cal A}],  
\end{equation}
where $x$ is the concentration of ${\cal A}$ element and $E[{\cal A}_x{\cal B}_{1-x}]$ is the free energy of composition ${\cal A}_x{\cal B}_{1-x}$. Also $E[{\cal B} ]$ and $E[{\cal A}]$ are the free energies of pure elements ${\cal A}$ and ${\cal B}$, respectively. Note that we assume here zero temperature where potential energy, configurational energy, and free energy are the same. At finite temperature values, however, one should differentiate when using these terminologies. 
  
Figure \ref{fig1} summarizes the results of the above strategy for Mg-Zn binary in HCP lattice. To reach convergence as described above, this process has demanded to gradually add and calculate the configurational energies of 402 unique structures from first-principles. The plot exhibits the formation energy of these 402 structures, calculated by Eq. (\ref{eq1}), as a function of $\rm Mg$ concentration. The energetically most favorable structures (compositions) are those with energies touching the convex hull (black curve). To make it more visible, we have marked those by solid red circles while other structures close to the convex hull are marked by green circles. 

We find that $\rm Zn$, $\rm MgZn_3$, $\rm Mg_{19}Zn_{45}$, $\rm MgZn$, $\rm Mg_{34}Zn_{30}$, and $\rm Mg$ construct the convex hull and are the most favorable compounds for HCP lattice. We also show four other compounds, i.e., $\rm Mg_{15}Zn_{49}$, $\rm Mg_{17}Zn_{47}$, $\rm Mg_{21}Zn_{43}$, and $\rm Mg_{3}Zn$ that are close to the convex hull. The free energies of these configurations are summarized in Table \ref{table1}. The corresponding real space \fr~configurations are displayed on the right hand side of Figure \ref{fig1}, and they show gradual changes in site occupations. The prominent feature of these structures is that they show a strong mixing of Mg and Zn in HCP lattice symmetry over the entire composition range. The corresponding $5\times 5\times 5$ extended supercells are presented in the Appendix, Fig. \ref{fig9}.

\begin{figure*}[t]
\begin{tabular}{cc}
\includegraphics[width=12.5cm,height=5.80cm]{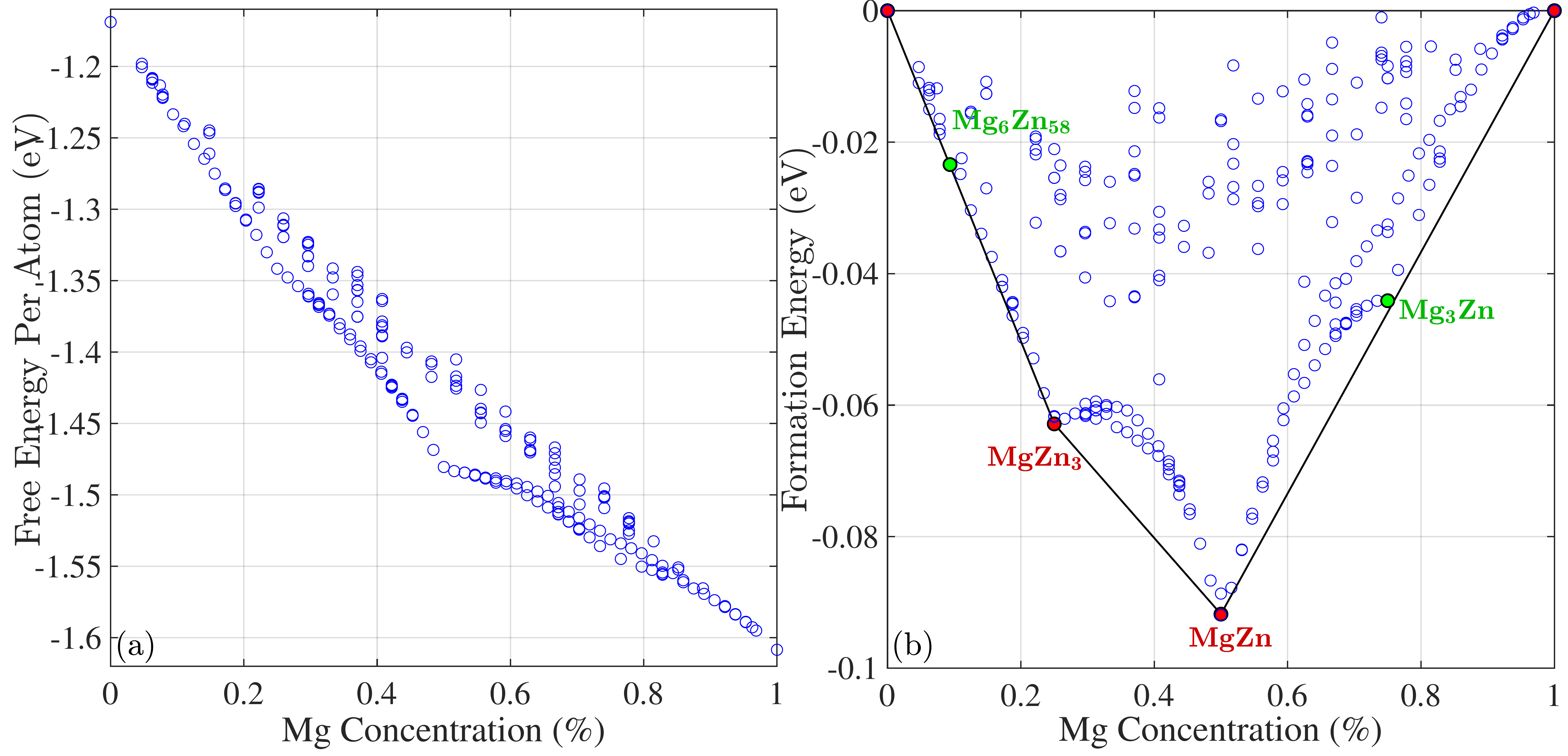} &
\includegraphics[trim=4.5cm 12.5cm 3.5cm 3.5cm, clip, width=5.5cm,height=5.80cm]{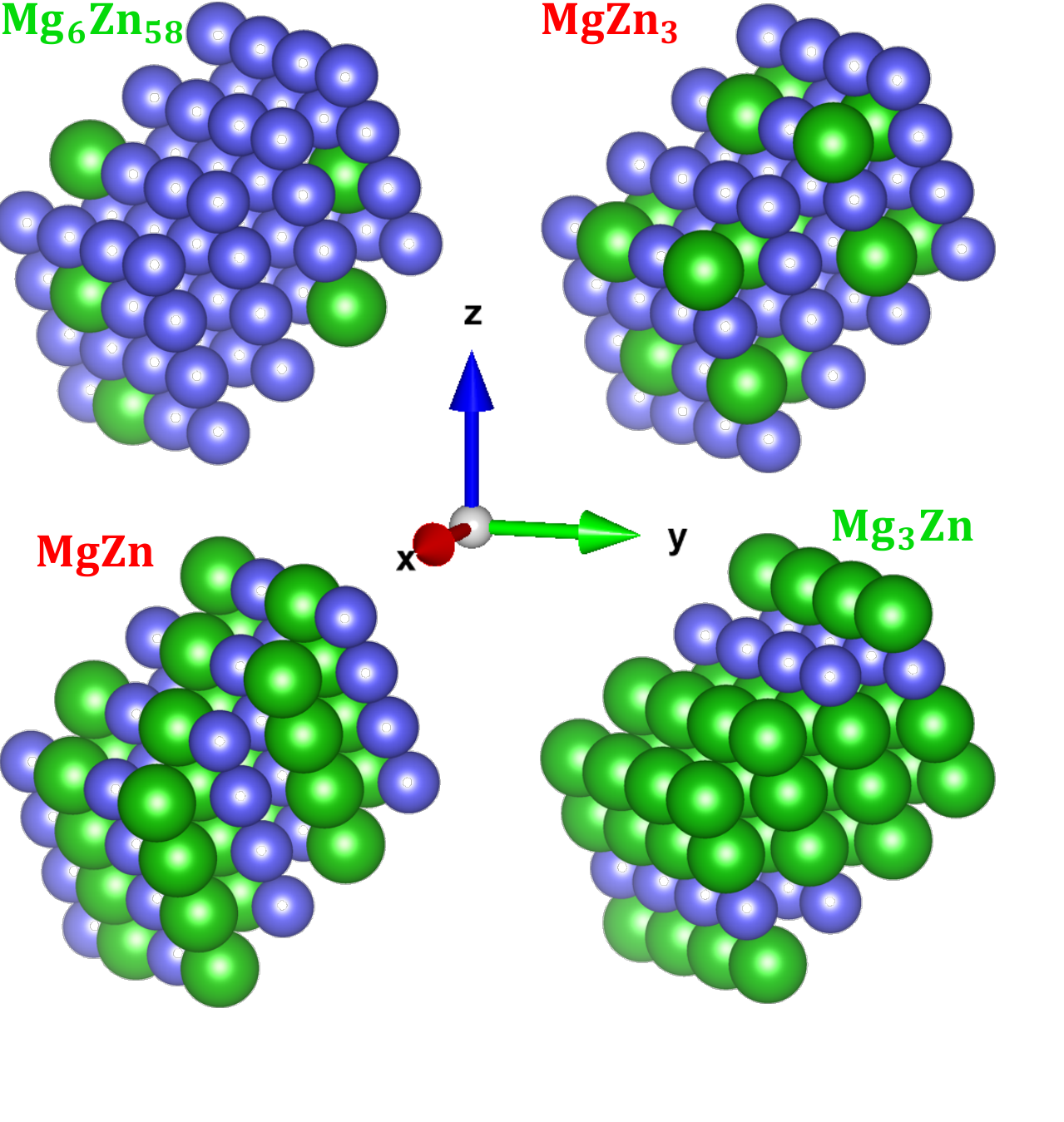} 
\end{tabular}
\caption{ (Color online). The Mg-Zn binary with the FCC lattice symmetry. (a) Normalized free energy of 232 configurations as a function of Mg concentration (in percentage $\%$). (b) The convex hull of these 232 unique configurations with \fr~unit cell structures contaning 64 atoms. The stable phases are $\rm MgZn_{3}$ and $\rm MgZn$ (marked by red), whereas $\rm Mg_{6}Zn_{58}$ and $\rm Mg_3Zn$ (marked by green) are close to the convex hull. The corresponding atomic structures are shown on the right side. The green and blue spheres stand for $\rm Mg$ and $\rm Zn$ atoms, respectively.  
 }
 \label{fig2}
\end{figure*}

Note that since we are restricted by \fr~structures due to computational limitations in DFT calculations, we are able to span the concentration range only by steps of $1.5625\%$. Therefore, there may exist other energetically favorable compounds close to the convex hull (namely, they fall into regions that are only explorable through larger atomic models). Nevertheless, the step of $1.5625\%$ is small enough to allow us to provide an overview of the convex hull for binary alloys. The stable phases with hexagonal symmetries in experiment are $\rm Mg_2Zn_3$, $\rm Mg_2Zn_{11}$, $\rm MgZn_2$, $\rm Mg_4Zn_7$, and $\rm MgZn$. \cite{A.Bendo,A.Singh,A.Lervik,T.F.Chung,W.Liu,H.Okamoto,J.Peng} The last case was also explored and confirmed in our results above. The ratios of the other two compositions, $1$:$2$ and $4$:$7$, do not match exactly with the \fr~structures considered in this study, but the predicted phase $\rm Mg_{21}Zn_{43}$ lends support also for these experimental observations. \cite{A.Bendo,A.Singh,A.Lervik,T.F.Chung,W.Liu,H.Okamoto,J.Peng} 

Another limitation considers the (possible) structural transition upon alloying or applying pressure. In this case, the searchable configuration space would considerably increases and include numerous new possibilities. We postpone addressing these interesting topics to our future works, and thus, assume here that no structural transitions beyond the HCP/FCC will take place under alloying and/or the application of pressure. Note that as close-packed structures the HCP and FCC lattices have the largest coordination number, i.e., 12 (sphere packing efficiency of $0.74$), and it does not seem plausible that other lattice structures (such as body-centered cubic) would emerge as a result of applying pressure. 

\begin{figure}[t!]
\includegraphics[width=8.5cm,height=10.10cm]{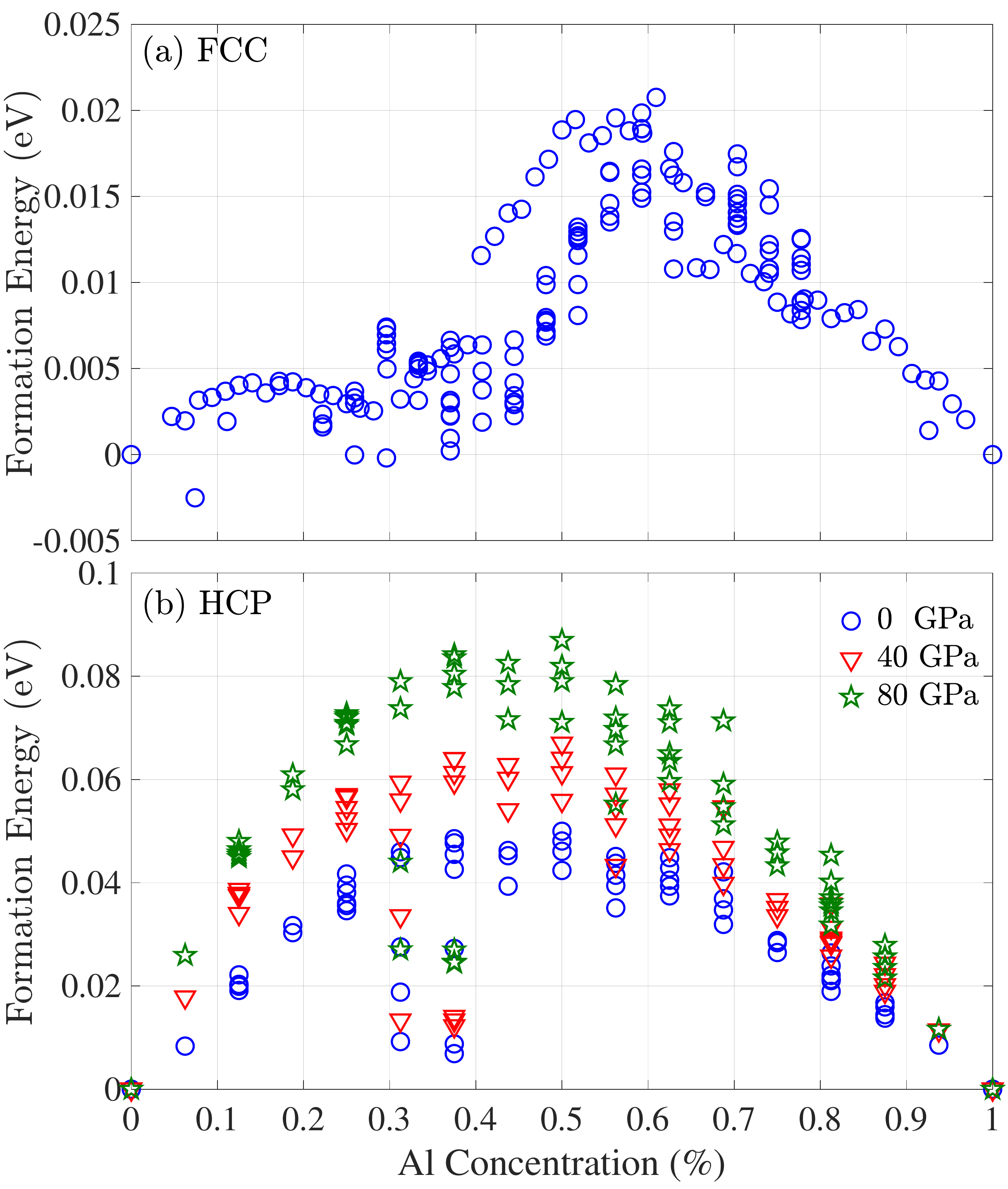} \caption{ (Color online). Formation energy of unique configurations of the Al-Zn binary as a function of Al concentration. (a) FCC lattice with $163$ unique configurations (3$\times$3$\times$3 and 4$\times$4$\times$4 structures). The external pressure is set zero. (b) HCP lattice and $64$ unique configurations under external pressure: P$=$$0$, $40$, $80$~GPa.   }
 \label{fig3}
\end{figure}
 
Next, we study the Mg-Zn binary with the FCC lattice symmetry. As bulk materials, $\rm Mg$ and $\rm Zn$ have the HCP lattice symmetry whereas $\rm Al$ exists as the FCC lattice symmetry. However, as it is fully confirmed by experiments, the solid solutes and clusters in $\rm Al$ dominated matrices tend to mimic the lattice symmetry of bulk $\rm Al$ and develop FCC-like sublattices. Therefore, in order to understand the properties of the Mg-Zn solid clusters in the $\rm Al$-dominated matrices one needs to study this binary with the FCC symmetry as well. Figure \ref{fig2}(a) shows the normalized free energy of 232 unique structures as a function of concentration. The FCC structures have been systemically collected in the DFT database as described above for the HCP case. Fig. \ref{fig2}(a) shows the normalized free energy and Figure \ref{fig2}(b) exhibits the associated formation energy obtained by Eq. (\ref{eq1}). Note that the HCP structures display exactly the same trend (not shown) as Fig. \ref{fig2}(a) as the energetical differences between the FCC and HCP configurations are very small. For the FCC lattice symmetry, the convex hull is made by $\rm Zn$, $\rm MgZn_3$ (by the green solid circles), $\rm MgZn$, and $\rm Mg$ marked by solid red circles. We have also marked $\rm Mg_6Zn_{58}$ and $\rm Mg_3Zn$ that have energies closer to the convex hull than other compounds. The normalized free energy of these compounds (shown by the blue circles) are summarized in Table \ref{table2}. The corresponding real space \fr~configurations are shown on the right hand side of Fig. \ref{fig2}. Interestingly, unlike for the HCP lattice, $\rm Mg_6Zn_{58}$ and $\rm MgZn_3$ suggest a full mixture of $\rm Mg$ and $\rm Zn$ atoms from $0\%$ to $30\%$ magnesium, while the stable $\rm MgZn$ phase and $\rm Mg_3Zn$ display that compositions containing more than $50\%$ of $\rm Mg$ tend to create layered structures. In order to ease the visibility of the mixture and layered compositions, we present viualizations of the $5\times 5\times 5$ extended cells in the Appendix, Fig. \ref{fig9}.     

\begin{figure*}[thp]
\includegraphics[width=18.0cm,height=7.80cm]{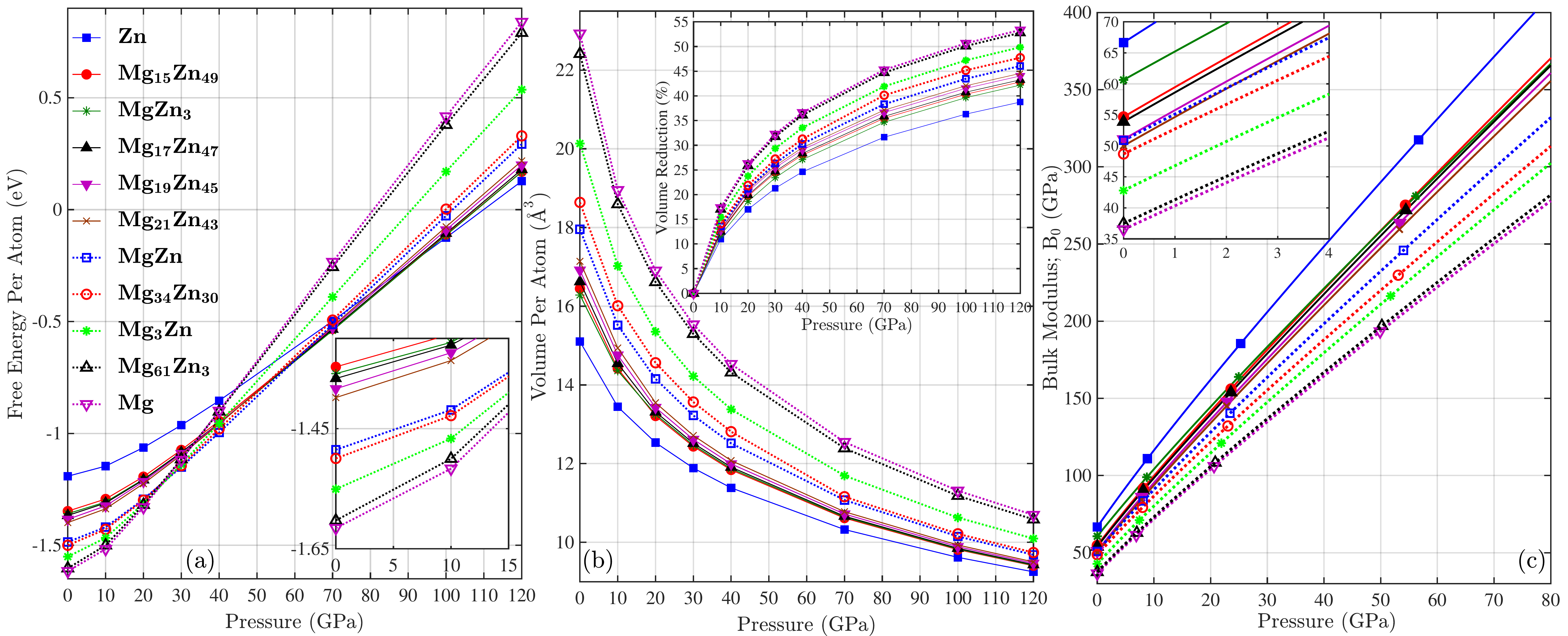} \\
\includegraphics[width=14.0cm,height=7.10cm]{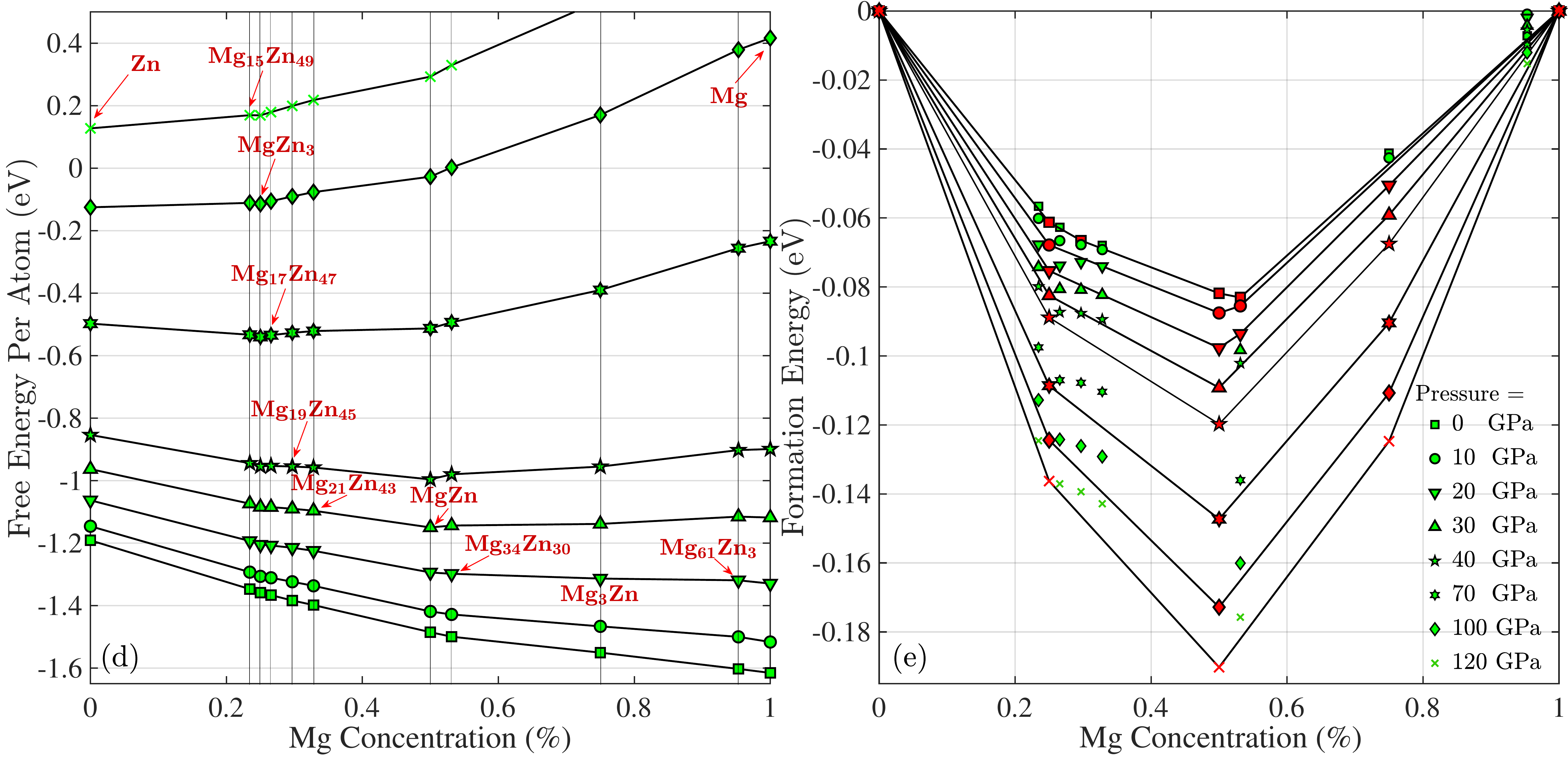}
\caption{ (Color online). The Mg-Zn binary with the HCP lattice symmetry.
(a) The free energy of marked phases in the convex hull, normalized by the number of atoms in supercell, as a function of applied isotropic pressure from $0$ GPa to $120$ GPa. (b) Change in the normalized volume of different compounds when applying the isotropic pressure. The inset panel shows the percentage of volume reduction due to the exerted pressure. (c) Bulk modulus, $\rm B_0$, of different compounds under pressure. The inset panel is a zoomed-in of the bulk modulus restricted to low pressures (from $0$ GPa to $5$ GPa). (d) Normalized free energy of different compositions as a function of Mg concentration (in percentage $\%$) at various values of the isotropic external pressure: P=$0, 10, 20, 30, 40, 70, 100, 120$ GPa. (e) Variation of convex hull and the evolution of the stable phases by increasing the external pressure. The stable phases are marked by red color, while the rest of compounds are shown by green.  
 }
 \label{fig4}
\end{figure*}

We have carried out the same computational procedure as above for the Al-Zn binary. Figure~\ref{fig3}(a) shows the formation energy as function of $\rm Al$ concentration for 164 unique structures, gradually collected through iterating the loop between the stages (i) and (v). The corresponding free energies are presented in Fig. \ref{fig10} in Appendix. Here we have considered both \tr~ and \fr~structures with the FCC lattice symmetry. The formation energy is either positive or takes very small negative values throughout the full composition range. The overall positive formation energy indicates that the Al-Zn binary tends to segregate always. This is confirmed by the atomic configurations of the lowest-energy structures (not shown) which show a strong segregation of $\rm Al$ and $\rm Zn$. To provide further insight for this binary, we show in Fig.~\ref{fig3}(b) the formation energy of the Al-Zn binary for the HCP lattice. Clearly, $\rm Al$ and $\rm Zn$ prefer to separate as for the FCC lattice symmetry. We also present results for the isotropic pressures of $40$~GPa and $80$~GPa in the HCP case. Introducing a non-zero pressure results in a stronger positive formation energy, and thus, it is unable to assist mixing of the $\rm Al$ and $\rm Zn$ elements. Note that, as pure Al and Zn belong to the FCC and HCP lattice symmetries, \cite{S.L.Chen} respectively, the final segregated compositions might be mixtures of these two symmetries. Nevertheless, in Al-rich and Zn-rich alloys, one can expect that the leading lattice symmetry is dictated by the FCC and HCP symmetries, respectively. 

Having determined the stable phases of Mg-Zn binary on the HCP and FCC lattices, we now subject these compounds (together with those nearby the convex hull) to the external pressure. Figure~\ref{fig4}(a) shows the normalized free energy of 11 compounds against externally applied isotropic pressure from $0$ to $120$~GPa. Under ambient conditions, bulk $\rm Mg$ and $\rm Zn$ have the lowest and highest free energy, respectively, and the other compounds are located between these two limits. By applying an external pressure, the free energy of all compounds increases and at high pressure regions, larger than $80$~GPa, the free energy becomes positive, limiting the formation of solid state compounds. Fig.~\ref{fig4}(a) illustrates that although $\rm Mg$-rich compounds are more stable at low pressure regime, they turn unstable faster than those with a higher concentration of $\rm Zn$ and there is a cross-over at $30$~GPa. Correspondingly, the $\rm Zn$-rich compounds show more stabilility under pressure, which can be used as a control knob for switching between the two types of alloys. Also, this finding can be expanded to large enough precipitates of these elements: By applying pressure to an Al matrix where $\rm Mg$ and $\rm Zn$ solute atoms form solid clusters, one may be able to externally control the type of the clusters to be either $\rm Mg$-rich or $\rm Zn$-rich. Figure~\ref{fig4}(b) shows the normalized volume of each compound against the applied isotropic pressure. Obviously, bulk $\rm Mg$ and $\rm Zn$ exhibit the highest and lowest volume reduction, respectively, and the rest are located between these two limits. The percentage of volume reduction confirms the higher stability of $\rm Zn$-rich compounds.  

\begin{figure*}[thp]
\includegraphics[width=18.0cm,height=7.80cm]{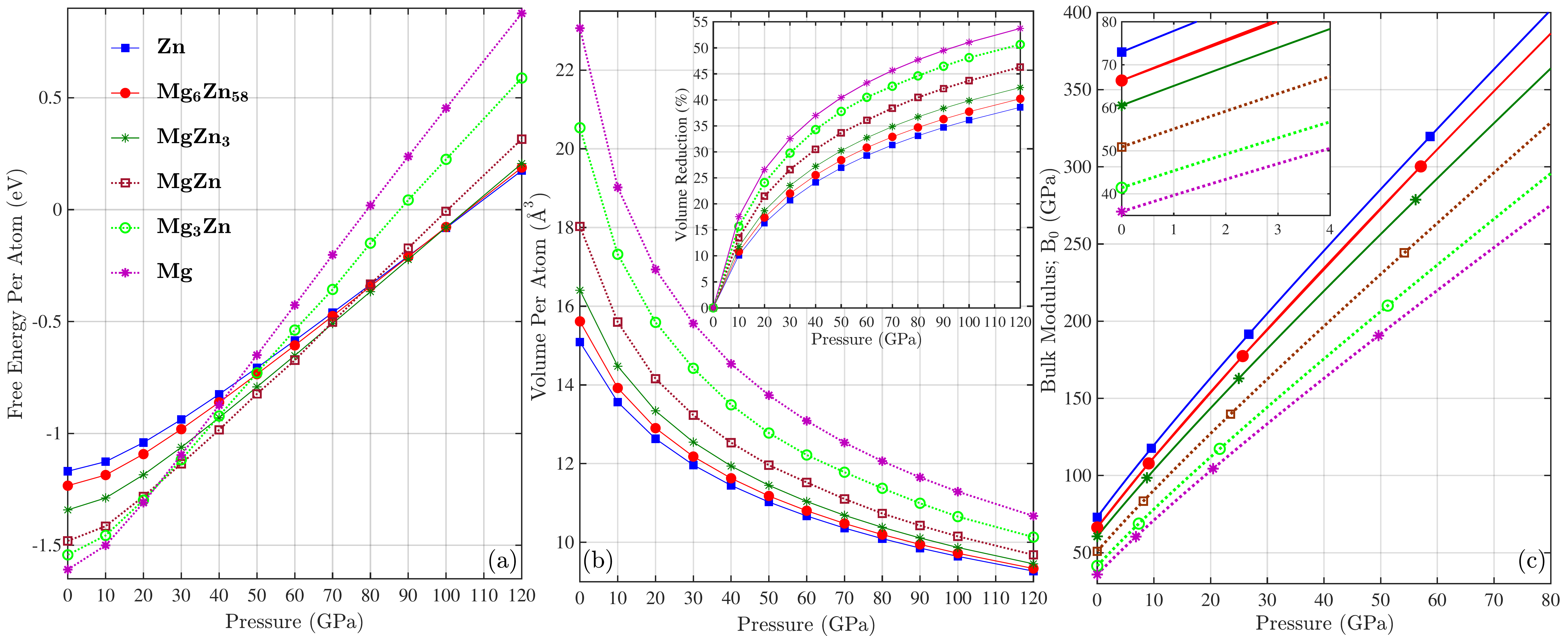} \\
\includegraphics[width=14.0cm,height=7.10cm]{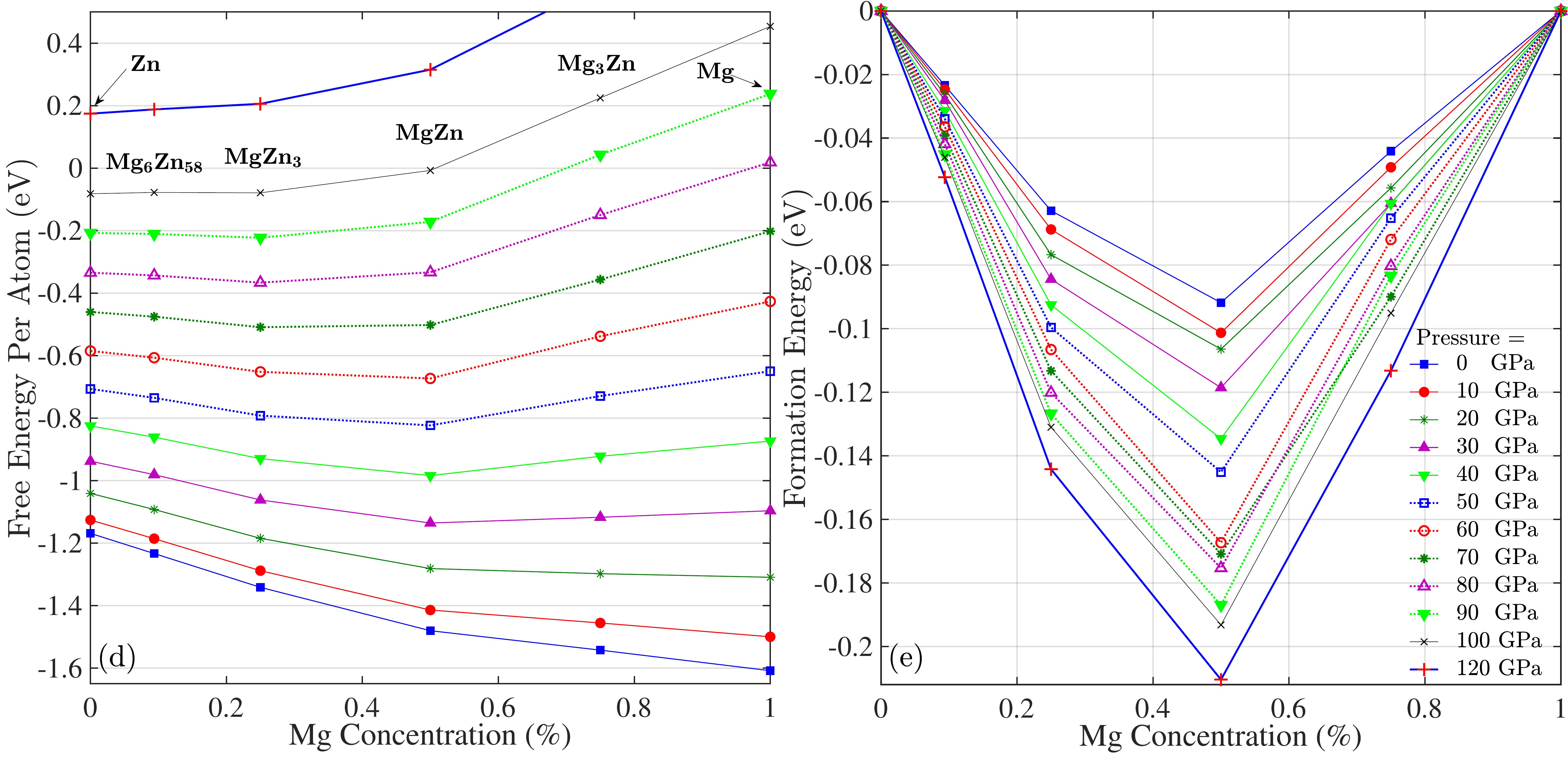}
\caption{ (Color online). The Mg-Zn binary with the FCC lattice symmetry. (a) Normalized free energy of the Mg-Zn compositions as a function of the externally applied isotropic pressure from $0$ GPa to $120$ GPa with a step of $10$ GPa. (b) Normalized volume of the compounds as a function of pressure. The inset panel exhibits the volume reduction of each compound as a function of pressure. (c) Bulk modulus as a function of pressure. The inset shows a zoomed-in shot of the bulk modulus at low pressures. (d) Normalized free energy as a function of Mg concentration (in percentage). (e) Formation energy of the Mg-Zn binary phases.
 }
 \label{fig5}
\end{figure*}

One important quantity that we can calculate now is the bulk modulus $B_0$ which measures the stiffness of a material against elastic deformation when subject to an external pressure. To this end, we make use of the Birch-Murnaghan empirical equation of state for pressure as a function of volume:
\begin{equation}\label{eq21}
P(V) = \frac{3B_0}{2}  \left\{  1 + \frac{3(B_0'-4)}{4} \left[ \left( \frac{V_0}{V} \right)^{ \frac{2}{3} } - 1\right ] \right\}    \left\{  \left( \frac{V_0}{V} \right)^{ \frac{7}{3} } -  \left( \frac{V_0}{V} \right)^{ \frac{5}{3} } \right\},
\end{equation}
in which $B_0$ is the bulk modulus, $B_0'$ is the derivative of the bulk modulus with respect to pressure, $V_0$ is the volume at zero pressure, $V$ is the volume, and $ P$ is the pressure. Also, the bulk modulus is closely linked to the speed of sound (mechanical waves) and the energy stored in a solid system, which is given by;
\begin{equation}\label{eq2}
B_0=-{V}\left( \frac{\partial P}{\partial V}\right).
\end{equation}
To evaluate Eq. (\ref{eq2}), we first find the set of parameter values that fits Eq. (\ref{eq21}) into our DFT data such that we obtain $P(V)$ numerically, and thus $B_0$ can be evaluated. Figure~\ref{fig4}(c) shows the bulk modulus of the Mg-Zn compounds. The calculated bulk moduli of the elemental $\rm Mg$ and $\rm Zn$ are in good agreement with those reported values in experiments. The calculated and measured $B_0$ are summarized in Table \ref{table3}. The bulk moduli of the Mg-Zn compounds are located between those of bulk $\rm Zn$ and $\rm Mg$ from top to bottom, respectively. Clearly, we can conclude that $\rm Zn$-rich compounds are stiffer than those with a high concentration of $\rm Mg$.

In order to evaluate the stability of these compounds subject to isotropic pressure, we show their normalized free energy values as a function of $\rm Mg$ concentration in Fig. \ref{fig4}(d) at several pressures. The associated convex hulls are presented in Fig.~\ref{fig4}(e). As evidenced above based on the bulk moduli, the free energy of bulk $\rm Mg$ (Fig.~\ref{fig4}(d)) has the largest variation. To make it clearer, we have marked stable phases, touching the convex hull, by solid red symbols, while those compounds away from the convex hull are displayed in green. We find that $\rm MgZn_3$ and $\rm MgZn$ show stability in the entire interval of the applied pressure. Interestingly, we see that $\rm Mg_{19}Zn_{45}$, which is a stable phase at zero pressure, becomes unstable at larger pressure values than $10$~GPa. Similarly, $\rm Mg_{34}Zn_{30}$ becomes unstable at pressures higher than $40$~GPa. Remarkably, $\rm Mg_3Zn$ becomes more stable at pressures higher than $20$~GPa as it not only touches the convex hull but also comprises a new minimum above $70$~GPa. 

Figure~\ref{fig5} is the FCC counterpart of Fig.~\ref{fig4} showing the effect of pressure on the stable phases $\rm Mg_6Zn_{58}$, $\rm MgZn_3$, $\rm MgZn$, and $\rm Mg_3Zn$. The normalized free energies reside between those of bulk $\rm Zn$ and $\rm Mg$ from zero pressure up to $30$~GPa. Similar to the HCP case, the free energy changes its sign above $80$~GPa, starting from bulk $\rm Mg$. At $30$~GPa, $\rm MgZn$ crosses with bulk $\rm Mg$ and becomes the lowest energy compound. At $80$~GPa, $\rm MgZn_3$ has the lowest free energy, and bulk $\rm Mg$ and bulk $\rm Zn$ change their places as the upper / lower boundary above $100$~GPa. Note that for both the HCP and FCC lattices, pressure-controlled alloying becomes possible around $30$~GPa, in particular at pressure values larger than $40$~GPa. Figure~\ref{fig5}(b) illustrates the normalized volume and the results are very similar to those for the HCP lattice in Fig.~\ref{fig4}(b). This concludes that $\rm Zn$-rich alloys are less sensitive to changes in pressure. For the bulk modulus (Fig.~\ref{fig5}(c)), as before, bulk $\rm Zn$ and bulk $\rm Mg$ exhibit the highest and lowest $B_0$ values, respectively. The bulk moduli of $\rm Mg$ on the HCP and FCC lattices are almost the same (Figs.~\ref{fig4}(c) and ~\ref{fig5}(c), insets). However, the $\rm Zn$ bulk modulus enhances by almost $15\%$ by the lattice change. Hence, our analysis demonstrates that $\rm Zn$-rich alloys with the FCC lattice symmetry are more favorable in terms of hardness and stability against external deformation.

\begin{figure}[t!]
\includegraphics[width=8.5cm,height=5.10cm]{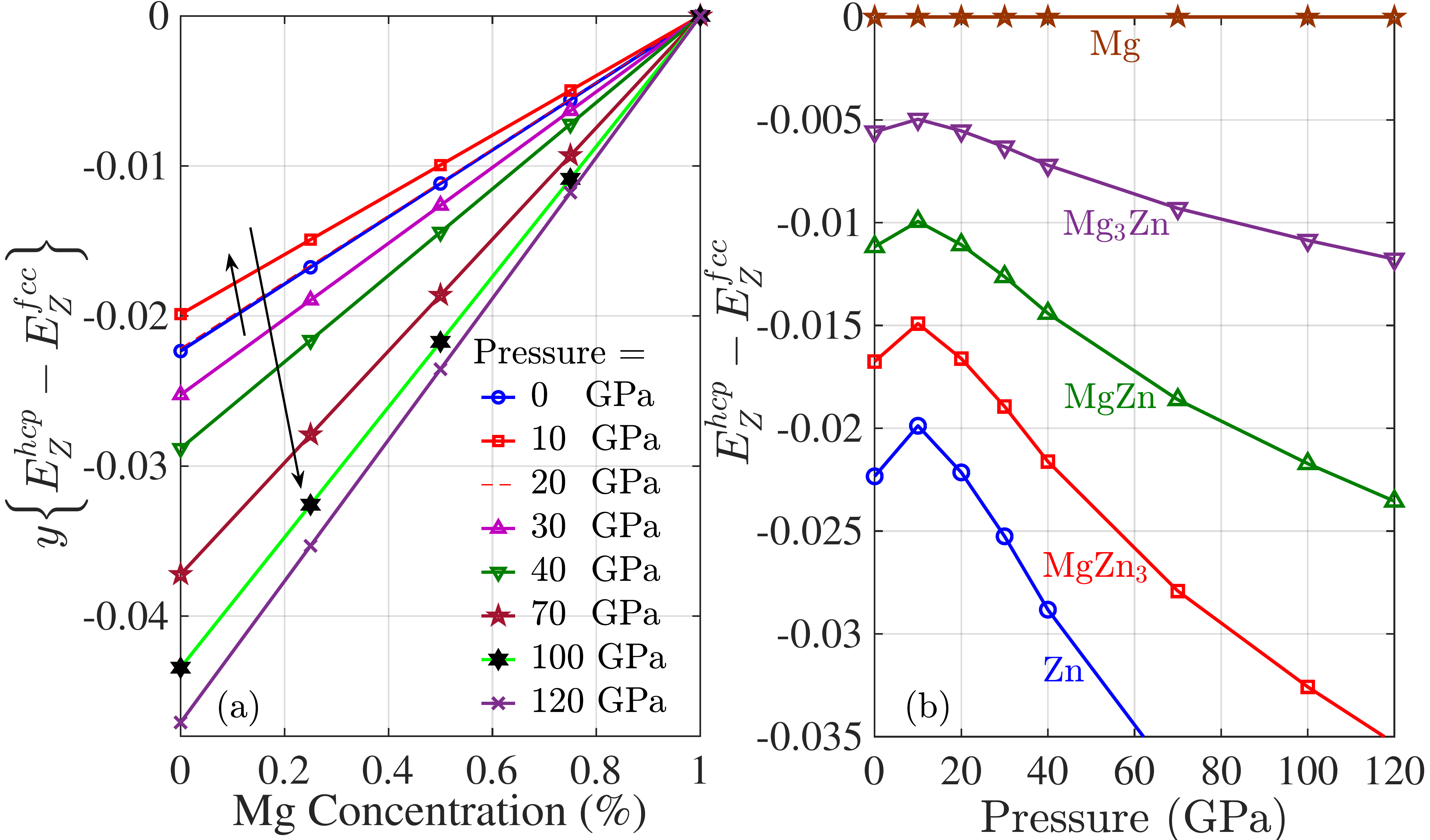} 
\caption{ (Color online). Structural energy difference of the $\rm Mg-Zn$ alloys under the isotropic external pressure. The vertical axis is in the units of energy (eV).  }
 \label{fig6}
\end{figure}

In Fig.~\ref{fig5}(d), we plot the normalized free energy as a function of $\rm Mg$ concentration and pressure and Figure~\ref{fig5}(e) displays the associated convex hull. Here, $\rm MgZn_3$, and $\rm MgZn$ are stable throughout the pressure interval $0$-$120$~GPa, while $\rm Mg_6Zn_{58}$ stays close to the convex hull. $\rm Mg_3Zn$ touches the convex hull above $20$~GPa but does not introduce a new mimimum. This differs from the HCP case where $\rm Mg_3Zn$ becomes stable as pressure increases. Therefore, we can conclude that $\rm Mg_3Zn$ with the FCC lattice symmetry has a limited stability when subject to pressure.        

Finally, let us consider a situation where the Mg-Zn binary can be a mixture of the HCP and FCC lattices. We calculate the mixing enthalpy and obtain the structural energy differences as follows;
\begin{subequations}
\begin{eqnarray}
&&\Delta H^\text{hcp}[\text{M}_x\text{Z}_y]  = E^\text{hcp}[\text{M}_x\text{Z}_y] - yE^\text{fcc}_\text{Z} - xE^\text{hcp}_\text{M}, \\
&&\Delta H^\text{fcc}[\text{M}_x\text{Z}_y]  = E^\text{fcc}[\text{M}_x\text{Z}_y] - yE^\text{hcp}_\text{Z} - xE^\text{fcc}_\text{M},
\end{eqnarray}
\end{subequations}
where $y\equiv 1-x$ and $E^\text{hcp,fcc}_{M,Z}$ are the energies of $\rm Mg$ and $\rm Zn$ on the HCP and FCC lattices. If we now add and subtract $yE^\text{hcp}_\text{Z}$ and $yE^\text{fcc}_\text{Z}$ on the right hand sides of the two equations, respectively, we find
\begin{subequations}
\begin{eqnarray}
&& \Delta H^\text{hcp}[\text{M}_x\text{Z}_y] = H^\text{hcp}[\text{M}_x\text{Z}_y] + y \Big\{ E^\text{hcp}_\text{Z} - E^\text{fcc}_\text{Z}\Big\},  \\
&& \Delta H^\text{fcc}[\text{M}_x\text{Z}_y] = H^\text{fcc}[\text{M}_x\text{Z}_y] + y \Big\{ E^\text{fcc}_\text{Z} - E^\text{hcp}_\text{Z}\Big\}, 
\end{eqnarray}
\end{subequations}
where $H^\text{hcp,fcc}[\text{M}_x\text{Z}_y]$ is the isostructural mixing enthalpy on the HCP and FCC lattices, given by Eq. (\ref{eq1}). Thus, the last terms in the above equations provide the contribution of the structural energy difference in the Mg-Zn alloys. We have plotted this quantity for $\rm Zn$, $\rm MgZn_3$, $\rm MgZn$, $\rm Mg_3Zn$, and $\rm Mg$ as a function of pressure in Fig. \ref{fig6}. The structural energy difference is very small, indicating that $\rm Mg$ and $\rm Zn$ can acquire FCC phase with a small energy cost. By increasing the pressure the structural energy difference first increases at $10$~GPa and then decreases while remaining negative the whole time, demonstrating that the HCP lattice remains always energetically preferrable. These findings can be confirmed by comparing the free energies in Tables \ref{table1} and \ref{table2}. We note here that according to the experimental observations, large enough precipitates develop hexagonal lattice symmetry under ambient conditions. \cite{A.Bendo,A.Singh,A.Lervik,T.F.Chung,W.Liu}

\begin{figure}[t!]
\includegraphics[width=8.5cm,height=8.10cm]{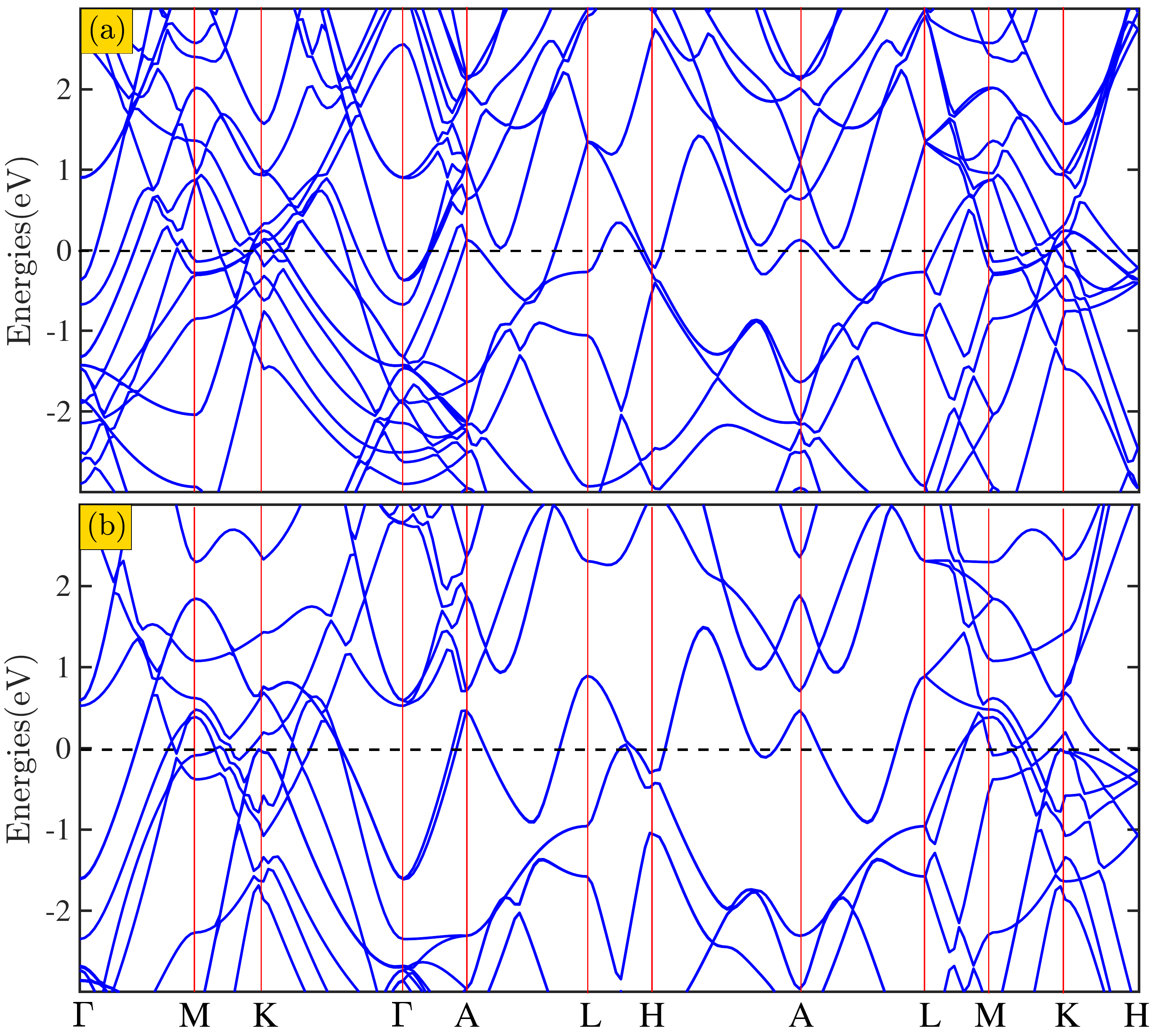} 
\caption{ (Color online). Electronic band structure of the $\rm MgZn$ with the HCP symmetry ($2 \times 2 \times 2$ unit cell) along the high-symmetry lines. The vertical lines exhibit the location of high-symmetry points. The Fermi level is shifted to zero energy. (a) Ambient condition ($P=0$ GPa); the normalized volume per atom is $18.3981 \AA$. (b) $P=120$ GPa; the normalized volume per atom is $9.8069 \AA$. 
}
 \label{fig7}
\end{figure}

\begin{figure*}[thp]
\includegraphics[width=18.0cm,height=9.70cm]{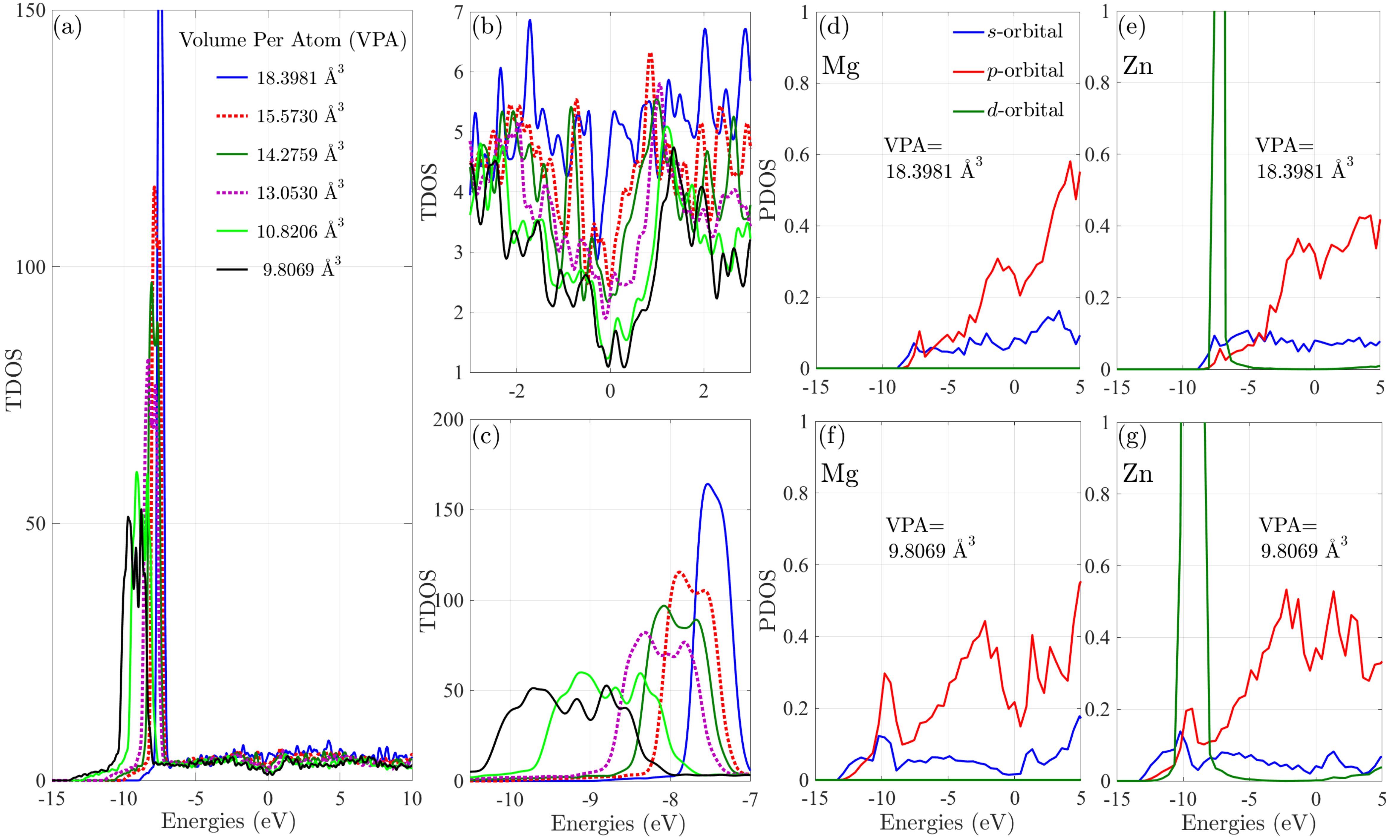} \\
\caption{ (Color online). Total and projected density of electronic states (TDOS/PDOS)  of $\rm MgZn$ with the HCP symmetry as a function of energy. The Fermi level is shifted to zero energy. TDOS is plotted for six values of volume per atom: $18.3981~\AA^3$ (0 GPa), $15.5730~\AA^3$ (10 GPa), $14.2759~\AA^3$ (20 GPa), $13.0530~\AA^3$ (30 GPa), and $9.8069~\AA^3$ (120 GPa). (a) TDOS in a broad range of energies: from $-15$~eV to $+10$~eV. (b)-(c) Zoomed-in TDOS at the Fermi level and the main peak in the valence band. (d)-(g) PDOS corresponding to the two extreme volumes $18.3981~\AA^3$ (0 GPa) and $9.8069~\AA^3$ (120 GPa).
 }
 \label{fig8}
\end{figure*}

\subsection{Electronic band structure and density of states}\label{dos}
DFT calculations include explicit information of the electronic structure for all configurations sampled in this study. Here, we present the density of states and band structure for the $\rm MgZn$ compound and HCP lattice symmetry, which is one of the lowest energy configurations located at the convex hull (see Fig. \ref{fig1}). In order to find out how the applied pressure can influence the electronic properties, we have systematically scaled down the unit cell volume and plotted the associated band structure, total density of states (TDOS), and projected density of states (PDOS). Figure \ref{fig7} illustrates the band structure along the high-symmetry lines. We have considered two different normalized volumes $18.3981\AA$ and $9.8069\AA$ in Figs. \ref{fig7}(a) and \ref{fig7}(b) corresponding to $P=0$ and $P=120$ GPa, respectively. The band energies are shifted such that the zero energy corresponds to the Fermi level. The effect of pressure is evident around the Fermi level where there are less bands at high pressure. Further, by selecting the symmetry points $\Gamma$ and $A$ as examples, one can see that there are bands which have been pushed at lower energies in the valence band.

The corresponding TDOS and PDOS for the $\rm MgZn$ compound are plotted in Fig. \ref{fig8}. TDOS (Fig. \ref{fig8}(a)) is shown as a function of the normalized volume per atom (VPA), corresponding to applied pressures between $P=0-120$ GPa. As seen, there is a prominent peak in the valence band above -10 eV which responds to pressure by broadening and shifting at lower energies (Fig. \ref{fig8}(c)). Analysis of the projections onto atomic $s$-, $p$-, $d$-orbitals (PDOS, Figs. \ref{fig8}d-g) reveals that this pressure-sensitive peak is associated with the $d$-electrons of Zn. Furthermore, TDOS at the Fermi level shows clear suppression with decreasing VPA, in accordance with the band structure in Fig. \ref{fig7}. Inspection of the corresponding region in PDOS indicates that this reduction is mainly due to reduction in the corresponding the $s-$ and $p$-weights of Mg. In general, PDOS shows broadening towards lower energies upon applying pressure.

\section{conclusions}\label{conclusions}
We have employed a configurational cluster expansion method, Markov chain Monte Carlo search algorithm, and first-principles computations in the framework DFT to determine the stable phases of the Mg-Zn and Al-Zn binaries on both the HCP and FCC lattice symmetries. In order to find the ground states, we have constructed several CE models, performed extensive searches in configuration space to systematically predict configurations close to ground state in the entire composition range, and carried out DFT calculations (optimizing cell and geometry) for predicted configurations. By tuning concentration with a step of $\rm 1.5625\%$ (64-atom system), we have found that $\rm MgZn_3$, $\rm Mg_{19}Zn_{45}$, $\rm MgZn$, and $\rm Mg_{34}Zn_{30}$ with the HCP symmetry, and $\rm MgZn_3$ and $\rm MgZn$ with FCC symmetry, are the stable phases of the relaxed Mg-Zn binary, while Al-Zn has a positive formation energy throughout the entire composition range for both the HCP and FCC lattices subject to zero/finite pressure. Our findings are in a good agreement with experimental observations where $\rm MgZn_2$, $\rm Mg_4Zn_7$, and $\rm MgZn$ compounds with hexagonal symmetry are found to be the most stable phases of larger precipitates. \cite{A.Bendo,A.Singh,A.Lervik,T.F.Chung,W.Liu} 

For the HCP lattice, the increase of external pressure to P$\approx$$10$~GPa results in that $\rm Mg_{19}Zn_{45}$ becomes unstable and $\rm Mg_{34}Zn_{30}$ is less favorable. At pressure values of the order of $20$~GPa, a new compound, $\rm Mg_{3}Zn$, shows stability and becomes more stable at higher pressure values, while $\rm Mg_{34}Zn_{30}$ turns unstable at pressure values larger than $\approx$$30$~GPa. For the FCC convex hull, the $\rm Mg_{3}Zn$ compound weakly touches the convex hull at P$\gtrsim$$20$~GPa, whereas the other compounds remain stable throughout the entire pressure range $0$-$120$~GPa. By visualizing the atomic structures of the stable compounds, we have found that the stable phases of the HCP alloys are mixtures of $\rm Mg$ and $\rm Zn$. Similarly, the FCC alloys with a high concentration of $\rm Zn$ are perfect mixtures of $\rm Mg$ and $\rm Zn$, whereas in the $\rm MgZn$ and $\rm Mg$-rich compounds the atomic configurations exhibit layered elemental ordering. Analysis of the bulk modulus, volume reduction, and free energy of these compounds under finite pressure shows that $\rm Zn$-rich compositions are more stable and have greater hardness than those with a high concentration of $\rm Mg$. 

Our results suggest that by applying an appropriate external pressure into Al matrices containing $\rm Mg$ and $\rm Zn$ solutes, one may be able to control the type of the early-stage precipitates; being either $\rm Zn$-rich or $\rm Mg$-rich. As the $\rm Zn$-rich solid clusters have large enough bulk modulus than those of the $\rm Mg$-rich ones, the external pressure at early stage clustering can control the overall hardness of the entire matrix and improve its resistance against external deformation.     

\acknowledgments
The DFT calculations were performed using the resources provided by UNINETT Sigma2 - the National Infrastructure for High Performance Computing and Data Storage in Norway. Financial support from the NTNU Digital Transformation program (Norway) for the project ALLDESIGN is gratefully acknowledged. M.A. thanks A. Lervik, E. Thronsen, and other expert colleagues in NTNU and SINTEF for several conversations.

\appendix

\section{First-principles data and extended unit cells}

Here, we present the numerical results from DFT calculations for the Mg-Zn compounds discussed in the main text. Table \ref{table1} contains the normalized free energy of the Mg-Zn compounds with the HCP lattice symmetry. Table \ref{table2} presents the normalized free energy of the Mg-Zn compunds with the FCC lattice symmetry, and table \ref{table3} summarizes the bulk moduli of several compositions on both the HCP and FCC lattices. We have also included the experimentally measured values for bulk moduli.   

To illustrate how the $\rm Mg$ and $\rm Zn$ elements are arranged in the high-symmetry compositions found in Figs. \ref{fig1} and \ref{fig2}, we have expanded the unit cells of $\rm MgZn_3$, $\rm MgZn$, and $\rm Mg_3Zn$ up to $5\times 5\times 5$ cells. We have rotated the cells for visualization purposes as the coordinate axis arrows indicate. 

Also, Fig. \ref{fig10} exhibits the DFT calculated free energies of predicted unique structures for the Al-Zn alloy both on FCC and HCP lattices. The formation energy counterparts are presented in Fig. \ref{fig3}. In Fig. \ref{fig10}(a), the free energy of the Al-Zn alloys with the FCC lattice symmetry is plotted as a function of Mg concentration. The external pressure is set zero. Figure \ref{fig10}(b) presents similar study except now the lattice is changed to HCP and the results of $40$~GPa and $80$~GPa pressure values in addition to $0$~GPa are shown. The results show that the application of an isotropic static pressure to the Al-Zn alloys decreases the free energy.

\onecolumngrid
\begin{table*}
\caption{ Free energy per atom (eV) and volume per atom ($\AA^3$) for the Mg-Zn compositions in the HCP lattice.}
 \label{table1}
\begin{tabular}{c |c c c c c c c c c c c}
\hline
\multicolumn{12}{c}{Free energy - HCP}\\
\hline
Pressure (GPa) & $\rm Zn$ & $\rm Mg_{15}Zn_{49}$ & $\rm MgZn_{3}$ & $\rm Mg_{17}Zn_{47}$ & $\rm Mg_{19}Zn_{45}$ & $\rm Mg_{21}Zn_{43}$ &  $\rm MgZn$ & $\rm Mg_{34}Zn_{30}$ & $\rm Mg_{3}Zn$ & $\rm Mg_{61}Zn_3$ & $\rm Mg$\\
\hline
0 & -1.1913 & -1.3473 & -1.3586 & -1.3667 & -1.3838 & -1.3984 & -1.4853  & -1.4997 & -1.5507 & -1.6028 & -1.6155\\
10 & -1.1461 & -1.2931 & -1.3066 & -1.3111 & -1.3239 & -1.3369 &  -1.4189 & -1.4285 & -1.4666 & -1.5002 & -1.5167\\
20 & -1.0632 & -1.1933 & -1.2051 & -1.2078 & -1.2151 & -1.2247 & -1.2941  & -1.2983 & -1.3137 & -1.3192 & -1.3297\\
30 & -0.9632 & -1.0738 & -1.0845 & -1.0850 & -1.0901 & -1.0964 & -1.1500  & -1.1439 & -1.1387 & -1.1154 & -1.1184\\
40 & -0.8539 & -0.9444 & -0.9541 & -0.9532 & -0.9550 & -0.9581 & -0.9962  & -0.9800 & -0.9553 & -0.9024 & -0.8991\\
70 & -0.4976 & -0.5333 & -0.5403 & -0.5346 & -0.5271 & -0.5214 & -0.5130  & -0.4935 & -0.3902 & -0.2562 & -0.2338\\
100 & -0.1254 & -0.1112 & -0.1143 & -0.1057 & -0.0906 & -0.0768 & -0.0273  & 0.0024 & 0.1702 & 0.3790 & 0.4164\\
120 & 0.1274 & 0.1695 & 0.1689 & 0.1793 & 0.1992 & 0.2179 &  0.2928 & 0.3296 & 0.5362 & 0.7901 & 0.8388\\
\hline
\hline
\multicolumn{12}{c}{Volume - HCP}\\
\hline
Pressure (GPa) & $\rm Zn$ & $\rm Mg_{15}Zn_{49}$ & $\rm MgZn_{3}$ & $\rm Mg_{17}Zn_{47}$ & $\rm Mg_{19}Zn_{45}$ & $\rm Mg_{21}Zn_{43}$ &  $\rm MgZn$ & $\rm Mg_{34}Zn_{30}$ & $\rm Mg_{3}Zn$ & $\rm Mg_{61}Zn_3$ & $\rm Mg$\\
\hline
0 & 15.1025 & 16.4484 & 16.2846 & 16.6239 & 16.9157 & 17.1406 &  17.9532 & 18.6384 & 20.1345 & 22.4132 & 22.9254\\
10 & 13.4443 & 14.4160 & 14.3613 & 14.5454 & 14.7575 & 14.9387 &  15.5188 & 16.0125 & 17.0235 & 18.5934 & 18.9490\\
20 & 12.5345 & 13.2108 & 13.2517 & 13.3135 & 13.4257 & 13.5469 & 14.1554  & 14.5631 & 15.3522 & 16.6138 & 16.9017\\
30 & 11.8872 & 12.4360 & 12.4712 & 12.5180 & 12.6144 & 12.7148 & 13.2239  & 13.5661 & 14.2218 & 15.2960 & 15.5360\\
40 & 11.3845 & 11.8403 & 11.8717 & 11.9114 & 11.9934 & 12.0783 &  12.5161 & 12.8124 & 13.3791 & 14.3178 & 14.5297\\
70 & 10.3258 & 10.6168 & 10.6393 & 10.6651 & 10.7195 & 10.7778 & 11.0751  & 11.1598 & 11.6948 & 12.3911 & 12.5478\\
100 & 9.6188 & 9.8141 & 9.8288 & 9.8492 & 9.8897 & 9.9324 & 10.1504  & 10.2173 & 10.6290 & 11.1834 & 11.3116\\
120 & 9.2505 & 9.4047 & 9.4157 & 9.4335 & 9.4671 & 9.5027 & 9.6834  & 9.7407 & 10.0954 & 10.5841 &10.6959 \\
\hline
\hline
\end{tabular}
\end{table*}

\begin{table*}
\caption{ Free energy per atom (eV) and volume per atom ($\AA^3$) for Mg-Zn compositions in the FCC lattice.}
 \label{table2}
\begin{tabular}{c |c c c c c c }
\hline
\multicolumn{7}{c}{Free energy - FCC}\\
\hline
Pressure (GPa) & $\rm Zn$ & $\rm Mg_{6}Zn_{58}$ & $\rm MgZn_{3}$ & $\rm MgZn$ & $\rm Mg_{3}Zn$ & $\rm Mg$\\
\hline
0    &  -1.1689 & -1.2336  & -1.3417  & -1.4805  & -1.5427  &  -1.6085 \\
10  & -1.1262  & -1.1859  & -1.2884  & -1.4144  & -1.4558  & -1.5000 \\
20  & -1.0410  & -1.0925  & -1.1848  & -1.2817  & -1.2980  & -1.3095 \\
30  & -0.9379  & -0.9810  & -1.0620  & -1.1358  & -1.1174  &  -1.0966\\
40  &  -0.8251 &  -0.8610 & -0.9298  &  -0.9840 & -0.9222  &  -0.8738\\
50  &  -0.7066 &  -0.7352 & -0.7920  &  -0.8234 & -0.7293  & -0.6500 \\
60  &  -0.5851 &  -0.6067 & -0.6520  & -0.6732  & -0.5381  &  -0.4266\\
70  & -0.4604  &  -0.4754 & -0.5090  & -0.5021  &  -0.3566 &  -0.2021\\
80  &  -0.3346 &  -0.3434 & -0.3664  & -0.3333  &  -0.1499 &  0.0187\\
90  &  -0.2073 &  -0.2107 &  -0.2228 & -0.1718  &   0.0430&  0.2375\\
100  & -0.0819  & -0.0778  & -0.0790  & -0.0073  &  0.2247 &  0.4537\\
120 & 0.1745  &  0.1880 & 0.2059  &  0.3153 & 0.5882 &  0.8770\\
\hline
\hline
\multicolumn{7}{c}{Volume - FCC}\\
\hline
Pressure (GPa) & $\rm Zn$ & $\rm Mg_{6}Zn_{58}$ & $\rm MgZn_{3}$ & $\rm MgZn$ & $\rm Mg_{3}Zn$ & $\rm Mg$\\
\hline
0    &  15.0892 & 15.6145  & 16.4053  & 18.0280  & 20.5353  &  23.0647 \\
10  & 13.5633  & 13.9217  & 14.4742  &  15.5968 & 17.3198  &  19.0178\\
20  & 12.6312  & 12.9019  & 13.3414  & 14.1572  & 15.5902  &  16.9342\\
30  & 11.9649  & 12.1799  & 12.5467  &  13.2325 & 14.4226  &  15.5586\\
40  &  11.4460 & 11.6277  &  11.9385 &  12.5335 & 13.4950  &  14.5354\\
50  &  11.0239 & 11.1780  &  11.4474 &  11.9608 & 12.7777  &  13.7380\\
60  &  10.6699 &  10.8030 &  11.0391 &  11.5233 &  12.2199 &  13.0870\\
70  &  10.3625 &  10.4792 & 10.6865  & 11.1031  &  11.7821 &  12.5341\\
80  &  10.0940 &  10.1971 &  10.3821 &  10.7341 &  11.3736 &  12.0630\\
90  &  9.8541 &  9.9471 & 10.1114  &  10.4269 &  10.9933 &  11.6510\\
100  & 9.6427  &  9.7231 &  9.8691 & 10.1523  &  10.6538 &  11.2869\\
120 &  9.2688 & 9.3355 &  9.4537 & 9.6819  & 10.1333  &  10.6698\\
\hline
\hline
\end{tabular}
\end{table*}

\begin{table*}
\caption{Calculated bulk moduli of the Mg-Zn alloys.}
 \label{table3}
\begin{tabular}{c |c c c c c c c c c c c}
\hline
\multicolumn{12}{c}{$B_0$ - HCP}\\
\hline
Pressure (GPa) & $\rm Zn$ & $\rm Mg_{15}Zn_{49}$ & $\rm MgZn_{3}$ & $\rm Mg_{17}Zn_{47}$ & $\rm Mg_{19}Zn_{45}$ & $\rm Mg_{21}Zn_{43}$ &  $\rm MgZn$ & $\rm Mg_{34}Zn_{30}$ & $\rm Mg_{3}Zn$ & $\rm Mg_{61}Zn_3$ & $\rm Mg$\\
\hline
0 & 66.6732 &  54.6725 & 60.6312  & 53.8911  & 51.0433  & 50.0149  & 50.8842  & 48.6929  &  42.8050 & 37.4733  &  36.5245 \\
\hline
\end{tabular}
\\
\begin{tabular}{c |c c c c c c }
\hline
\multicolumn{7}{c}{$B_0$ - FCC}\\
\hline
Pressure (GPa) & $\rm Zn$ & $\rm Mg_{6}Zn_{58}$ & $\rm MgZn_{3}$ & $\rm MgZn$ & $\rm Mg_{3}Zn$ & $\rm Mg$\\
\hline
0    &  72.9948 & 66.3637  & 60.5714  & 50.9126  & 41.4331  &  35.8843 \\
\hline
\end{tabular}
\\
\begin{tabular}{c |c c c c c c }
\hline
\multicolumn{6}{c}{$B_0$ - Experiment}\\
\hline
Pressure (GPa) & $\rm Zn$ & $\rm MgZn_{3}$ & $\rm MgZn$ & $\rm Mg_{3}Zn$ & $\rm Mg$\\
\hline
0    &  70.0 & -  & -  & -  & 45.0 \\
\hline
\hline
\end{tabular}
\end{table*}

\twocolumngrid

\begin{figure*}[t]
\includegraphics[trim=0.0cm 13.0cm 0.0cm 3.8cm, clip, width=17.50cm,height=10.0cm]{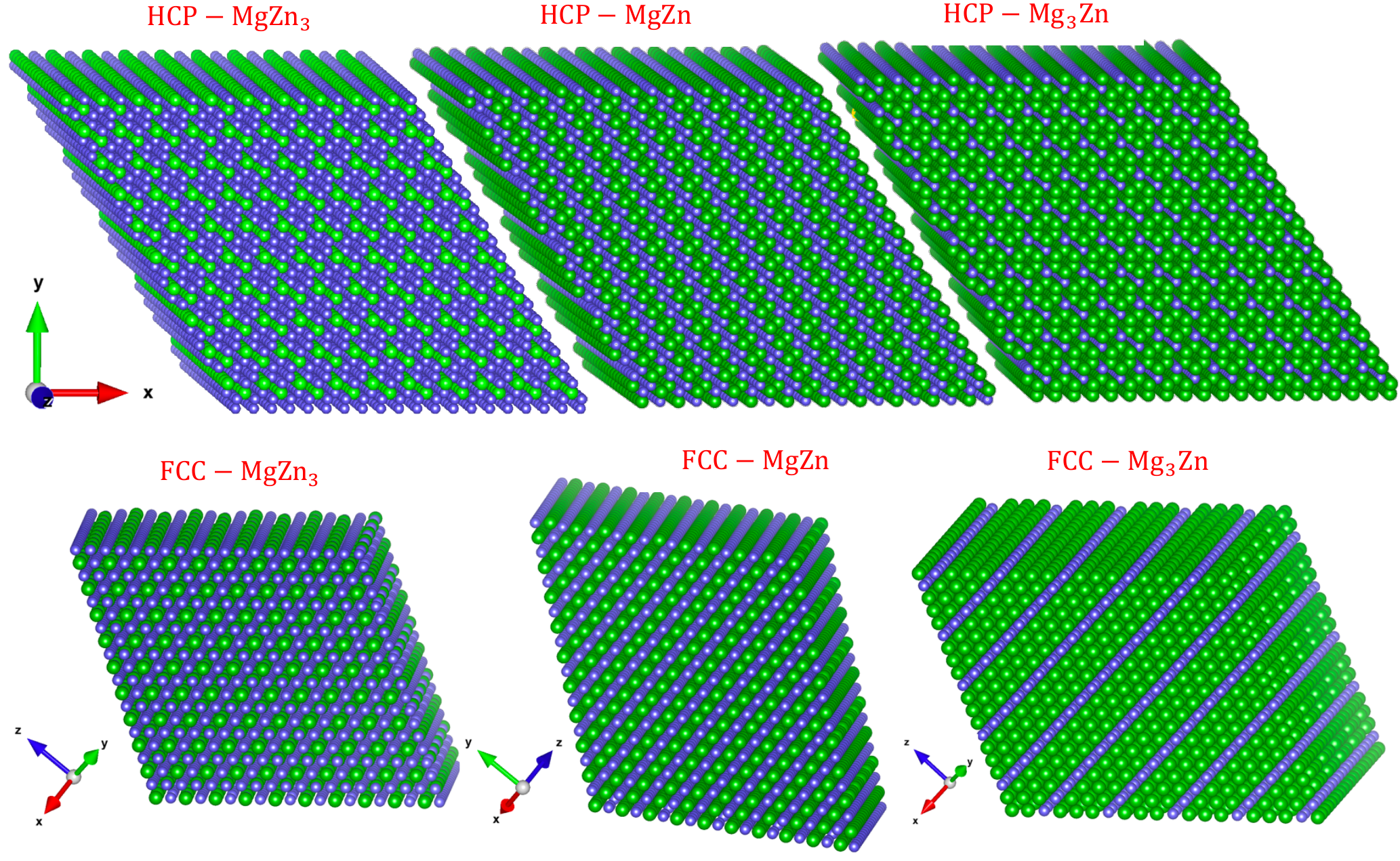} 
\caption{ (Color online). The high-symmetry HCP and FCC compositions of the Mg-Zn binary shown in the convex hull plots in the main text. The unit cells are replicated $5\times 5\times 5$ times.   
 }
 \label{fig9}
\end{figure*}

\begin{figure}[t!]
\includegraphics[width=8.5cm,height=10.10cm]{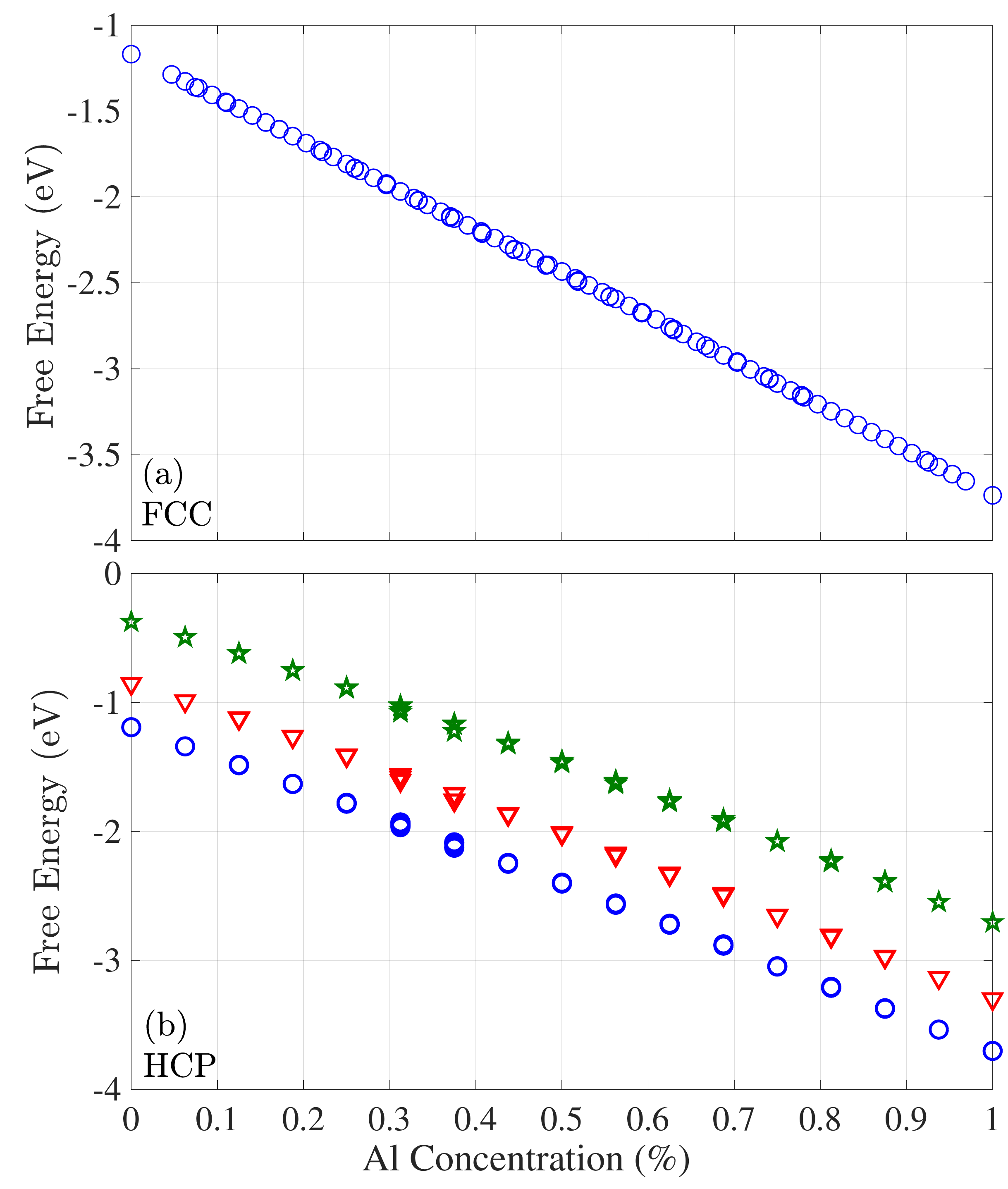} \caption{ (Color online). Free energy of unique configurations of the Al-Zn binary as a function of Al concentration. (a) FCC lattice with $163$ unique configurations (3$\times$3$\times$3 and 4$\times$4$\times$4 structures). The external pressure is set zero. (b) The HCP lattice and $64$ unique configurations under external pressure: P$=$$0$~GPa (circles), $40$~GPa (triangles), $80$~GPa (astricks).   }
 \label{fig10}
\end{figure}


\begin{thebibliography}{99}

\bibitem{F.W.Gayle} F.W. Gayle, M. Goodway, {\it Precipitation hardening in the first aerospace aluminum alloy: the wright flyer crankcase}, \href{https://science.sciencemag.org/content/266/5187/1015}{Science 266, 1015 (1994)}.

\bibitem{T.M.Pollock} T.M. Pollock, {\it Weight loss with magnesium alloys}, \href{https://science.sciencemag.org/content/328/5981/986.short}{Science 328,
986 (2010)}.

\bibitem{J.F.Nie} J.F. Nie, {\it Precipitation and hardening in magnesium alloys}, \href{https://link.springer.com/article/10.1007/s11661-012-1217-2}{Metall. Mater.
Trans. A 43, 3891 (2012)}.

\bibitem{Z.Wu} Z. Wu, M. F. Francis and W. A. Curtin, {\it Magnesium interatomic potential for simulating plasticity and fracture phenomena}, \href{https://iopscience.iop.org/article/10.1088/0965-0393/23/1/015004/meta}{Modelling Simul. Mater. Sci. Eng. 23, 015004 (2015)}.

\bibitem{Y.-M.Kim} Y.-M. Kim, N. J. Kim, B.-J. Lee , {\it Atomistic Modeling of pure $\rm Mg$ and $\rm Mg–Al$ systems}, \href{https://www.sciencedirect.com/science/article/pii/S0364591609000583}{CALPHAD: Computer Coupling of Phase Diagrams and Thermochemistry 33, 650 (2009)}.

\bibitem{P.Mao} P. Mao, B. Yu, Z. Liu, F. Wang, Y. Ju, {\it First-principles calculations of structural, elastic and electronic properties of $\rm AB_2$\rm  type intermetallics in $\rm Mg–Zn–Ca–Cu$ alloy}, \href{https://www.sciencedirect.com/science/article/pii/S2213956713000431}{Journal of Magnesium and Alloys 1, 256 (2013)}.

\bibitem{D.E.Dickel} D. E. Dickel, M. Baskes, I. Aslam and C. D. Barrett, {\it New interatomic potential for $\rm Mg–Al–Zn$ alloys with specific application to dilute Mg-based alloys}, \href{https://iopscience.iop.org/article/10.1088/1361-651X/aabaad/meta}{Modelling Simul. Mater. Sci. Eng. 26, 045010 (2018)}.

\bibitem{B.Zou} B. Zou, Z.Q. Chen, C.H. Liu, J.H. Chen, {\it Vacancy-$\rm Mg$ complexes and their
evolution in early stages of aging of $\rm Al–Mg$ based alloys}, \href{https://www.sciencedirect.com/science/article/pii/S0169433214001184}{Appl. Surf. Sci. 298, 50 (2014)}.

\bibitem{davidkl} D. Kleiven a, O. L. {\O}degard, K. Laasonen, J. Akola, {\it Atomistic simulations of early stage clusters in $\rm Al–Mg$ alloys}, \href{https://www.sciencedirect.com/science/article/pii/S1359645418310012}{Acta Materialia 166, 484 (2019)}.

\bibitem{B.Zheng} B. Zheng, L. Zhaoa, X.B. Hua, S.J. Donga, H. Li, {\it First-principles studies of $\rm Mg_{17}Al_{12}, Mg_{2}Al_3, Mg_2Sn, MgZn_2, Mg_2Ni, Al_3Ni$ phases}, \href{https://www.sciencedirect.com/science/article/pii/S0921452618307853}{Physica B: Condensed Matter 560, 255 (2019)}.

\bibitem{T.Tsurua} T. Tsurua, M. Yamaguchi, K. Ebihara, M. Itakura, Y. Shiihara, K. Matsuda, H. Toda, {\it First-principles study of hydrogen segregation at the $\rm MgZn_2$ precipitate in $\rm Al–Mg–Zn$ alloys}, \href{https://www.sciencedirect.com/science/article/pii/S0927025618301563}{Computational Materials Science 148, 301 (2018)}.

\bibitem{C.Ravi} C. Ravi, C. Wolverton, {\it First-principles study of crystal structure and stability
of $\rm AlMgSi(Cu)$ precipitates}, \href{https://www.sciencedirect.com/science/article/pii/S1359645404003106}{Acta Mater. 52, 4213 (2004)}.

\bibitem{A.V.Ruban} A.V. Ruban, I.A. Abrikosov, {\it Configurational thermodynamics of alloys from
first principles: effective cluster interactions}, \href{https://iopscience.iop.org/article/10.1088/0034-4885/71/4/046501/meta}{Rep. Prog. Phys. 71, 046501 (2008)}.

\bibitem{A.Yu.Stroev} A.Yu. Stroev, O.I. Gorbatov, Yu.N. Gornostyrev, P.A. Korzhavyi, {\it Solid solution decomposition and Guinier-Preston zone formation in $\rm AlCu$ alloys: a kinetic theory with anisotropic interactions}, \href{https://journals.aps.org/prmaterials/abstract/10.1103/PhysRevMaterials.2.033603}{Phys. Rev. Mater. 2, 033603 (2018)}.

\bibitem{H.S.Jang} H.-S. Jang, K.-M. Kim, B.-J. Lee, {\it Modified embedded-atom method interatomic potentials for pure $\rm Zn$ and $\rm Mg–Zn$ binary system}, \href{https://www.sciencedirect.com/science/article/pii/S0364591617302055}{Calphad 60, 200 (2018)}.

\bibitem{M.I.Baskes} M. I. Baskes, {\it Commentary on 'modified embedded atom method potentials for hcp metals' {\rm M. I. Baskes and R. A. Johnson (1994)} Modelling Simul. Mater. Sci. Eng.—the early basis for modeling hcp materials using MEAM}, \href{https://iopscience.iop.org/article/10.1088/1361-651X/aa817e/meta}{Modelling Simul. Mater. Sci. Eng. 25 071002 (2017)}.

\bibitem{J.M.Sanchez} J.M. Sanchez, {\it Cluster expansion and the configurational theory of alloys}, \href{https://journals.aps.org/prb/abstract/10.1103/PhysRevB.81.224202}{Phys. Rev. B 81, 224202 (2010)}.

\bibitem{S.Wang} S. Wang, Y. Zhao , H. Guo, F. Lan and H. Hou, {\it Mechanical and Thermal Conductivity Properties of Enhanced Phases in $\rm Mg–Zn–Zr$ System from First Principles}, \href{ https://www.mdpi.com/1996-1944/11/10/2010}{ Materials 11, 2010 (2018)}.

\bibitem{Y.Z.Ji} Y.Z. Ji, A. Issa, T.W. Heo, J.E. Saal, C. Wolverton, L.Q. Chen, {\it Predicting $\beta$' precipitate morphology and evolution in $\rm Mg–RE$ alloys using a combination of first-principles calculations and phase-field modeling}, \href{https://www.sciencedirect.com/science/article/pii/S1359645414003401}{Acta Mater. 76, 259 (2014)}.

\bibitem{T.E.M.Staab} T.E.M. Staab, B. Klobes, I. Kohlbach, B. Korff, M. Haaks, E. Dudzik, K. Maier, {\it Atomic structure of pre-Guinier-Preston and Guinier-Preston-Bagaryatsky
zones in {\rm Al}-alloys}, \href{https://iopscience.iop.org/article/10.1088/1742-6596/265/1/012018}{J. Phys. Conf. Ser. 265, 012018 (2011)}.

\bibitem{S.R.Nayak} S.R. Nayak, C.J. Hung, R.J. Hebert, S.P. Alpay, {\it Atomistic origins of Guinier-Preston zone formation and morphology in $\rm Al–Cu$ and $\rm Al–Ag$ alloys from first principles}, \href{https://www.sciencedirect.com/science/article/pii/S1359646218307127}{Scr. Mater. 162, 235 (2019)}.

\bibitem{G.Yi} G. Yi, W. Zeng, J.D. Poplawsky, D.A. Cullen, Z. Wang, M.L. Free, {\it Characterizing
and modeling the precipitation of $\rm Mg$-rich phases in $\rm Al 5xxx$ alloys aged at
low temperatures}, \href{https://www.sciencedirect.com/science/article/pii/S1005030217300506}{J. Mater. Sci. Technol. 33, 991 (2017)}.

\bibitem{M.Liu} M. Liu, H. Fu, L. Tian, W. Xiao, Q. Peng, C. Ma, {\it Nucleation and growth mechanisms of nano-scaled $\rm Si$ precipitates in $\rm Al–7Si$ supersaturated solid
solution}, \href{https://www.sciencedirect.com/science/article/pii/S0264127517302174}{Mater. Des. 121, 373 (2017)}.

\bibitem{C.Wang} C. Wang, T.-L. Huang, H.-Y. Wang, X.-N. Xue, Q.-C. Jiang, {\it Effects of distributions of $\rm Al$, $\rm Zn$ and $\rm Al+Zn$ atoms on the strengthening potency of $\rm Mg$ alloys: A first-principles calculations}, \href{https://www.sciencedirect.com/science/article/pii/S0927025615002165}{Computational Materials Science 104, 23 (2015)}.

\bibitem{G.Esteban-Manzanares} G. Esteban-Manzanares, A. Ma, I. Papadimitriou, E. Martínez and J. LLorca, {\it Basal dislocation/precipitate interactions in $\rm Mg–Al$ alloys: an atomistic investigation}, \href{https://iopscience.iop.org/article/10.1088/1361-651X/ab2de0/meta}{Modelling Simul. Mater. Sci. Eng. 27 075003 (2019)}.



\bibitem{T.Saito} T. Saito, E.A. Mørtsell, S. Wenner, C.D. Marioara, S.J. Andersen, J. Friis,
K. Matsuda, R. Holmestad, {\it Atomic structures of precipitates in $\rm Al–Mg–Si$ alloys with small additions of other elements}, \href{https://onlinelibrary.wiley.com/doi/abs/10.1002/adem.201800125}{Adv. Eng. Mater. 20, 1800125 (2018)}.

\bibitem{S.J.Andersen} S.J. Andersen, C.D. Marioara, J. Friis, S. Wenner, R. Holmestad, {\it Precipitates in
aluminium alloys}, \href{https://www.tandfonline.com/doi/abs/10.1080/23746149.2018.1479984}{Adv. Phys. X 3, 1479984 (2018)}.

\bibitem{S.Wenner} T. Saito, S. Wenner, E. Osmundsen, C. D. Marioara, S. J. Andersen, J. R{\o}yset, W. Lefebvre and R. Holmestad, {\it The effect of Zn on precipitation in $\rm Al–Mg–Si$ alloys}, \href{https://www.tandfonline.com/doi/full/10.1080/14786435.2014.913819}{Philos. Mag. 94, 2410 (2014)}.

\bibitem{W.Chrominski} W. Chrominski, S. Wenner, C. D. Marioara, R. Holmestad, M. Lewandowska, {\it Strengthening mechanisms in ultrafine grained $\rm Al–Mg–Si$ alloy processed by hydrostatic extrusion–Influence of ageing temperature}, \href{https://www.sciencedirect.com/science/article/pii/S0921509316306360}{Mat. Sci. and Eng. 669, 447 (2016)}.

\bibitem{E.Christiansen} E. Christiansen, C. D. Marioara, K. Marthinsen, O. S. Hopperstad, R. Holmestad, {\it Lattice rotations in precipitate free zones in an $\rm Al–Mg–Si$ alloy}, \href{https://www.sciencedirect.com/science/article/pii/S1044580318303735}{Mat. Charac. 144, 522 (2018)}.

\bibitem{Y.Xin} Y. Xin, X. Zhou, H. Chen, J.-F. Nie, H. Zhang, Y. Zhang, Q. Liu, {\it Annealing hardening in detwinning deformation of $\rm Mg–3Al–1Zn$ alloy}, \href{https://www.sciencedirect.com/science/article/pii/S0921509313013178}{Materials Science and Engineering A 594, 287 (2014)}.

\bibitem{W.Lefebvre} W. Lefebvre, V. Kopp, C. Pareige, {\it Nano-precipitates made of atomic pillars revealed by single atom detection in a {\rm Mg–Nd} alloy}, \href{https://aip.scitation.org/doi/abs/10.1063/1.3701272}{Appl. Phys. Lett. 100, 141906 (2012)}.

\bibitem{X.Gao} X. Gao, J.F. Nie, {\it Enhanced precipitation-hardening in $\rm Mg–Gd$ alloys containing $\rm Ag$ and $\rm Zn$}, \href{https://www.sciencedirect.com/science/article/pii/S1359646207008342}{Scripta Mater. 58, 619 (2008)}.

\bibitem{J.F.Nie2} J.F. Nie, N.C. Wilson, Y.M. Zhu, Z. Xu, {\it Solute clusters and GP zones in binary
$\rm Mg–Re$ alloys}, \href{https://www.sciencedirect.com/science/article/pii/S1359645415301531}{Acta Mater. 106, 260 (2016)}.

\bibitem{H.D.Zhao} H. D. Zhao, G. W. Qin, Y. P. Ren, W. L. Pei, D. Chen, Y. Guo, {\it The maximum solubility of $\rm Y$ in $\alpha$-$\rm Mg$ and composition ranges of $\rm Mg_{24}Y_{5-x}$ and $\rm Mg_{2}Y_{1-x}$ intermetallic phases in {\rm Mg–Y} binary system}, \href{https://www.sciencedirect.com/science/article/pii/S0925838810023479}{J. Alloys Compd 509, 627 (2011)}.

\bibitem{W.Sun} W. Sun, Y. Zhu, R. Marceau, L. Wang, Q. Zhang, X. Gao, C. Hutchinson, {\it Precipitation strengthening of aluminum alloys by room-temperature cyclic plasticity}, \href{https://science.sciencemag.org/content/363/6430/972.abstract}{10.1126/science.aav7086}.

\bibitem{A.Bendo} A. Bendo, T. Maeda, K. Matsuda, A. Lervik, R. Holmestad, C. D. Marioara, K. Nishimura, N. Nunomura,
H. Toda, M. Yamaguchi, K.-i. Ikeda and T. Homma, {\it Characterisation of structural similarities of
precipitates in $\rm Mg–Zn$ and $\rm Al–Zn–Mg$ alloys
systems}, \href{https://www.tandfonline.com/doi/figure/10.1080/14786435.2019.1637032?scroll=top&needAccess=true}{https://doi.org/10.1080/14786435.2019.1637032}.

\bibitem{A.Lervik} A. Lervik, C. D. Marioara, M. Kadanik, J. C. Walmsley, B. Milkereit, and R. Holmestad, {\it Precipitation in an extruded {\rm AA7003} aluminium alloy: observations of {\rm 6xxx}-type hardening phases }, \href{??}{to be published}.

\bibitem{H.Okamoto} H. Okamoto, {\it Comment on Mg-Zn (magnesium-zinc)}, \href{https://pascal-francis.inist.fr/vibad/index.php?action=getRecordDetail&idt=4202941}{J. Phase Equilibria 15, 129 (1994)}.

\bibitem{J.Peng} J. Peng, L .Zhong, Y. Wang, Y. Lu, F. Pan, {\it Effect of extrusion temperature on the microstructure and thermal conductivity of $\rm Mg_{2.0} Zn_{1.0} Mn_{0.2} Ce$ alloys}, \href{https://www.sciencedirect.com/science/article/pii/S0264127515302902}{Materials and Design 87, 914 (2015)}.

\bibitem{A.Singh} A. Singh and J. Rosalie. {\it Lattice correspondence and growth structures of monoclinic ${\rm Mg_4Zn_7}$ phase growing on an icosahedral quasicrystal }, \href{https://www.mdpi.com/2073-4352/8/5/194}{Crystals 8, 194 (2018)}.

\bibitem{T.F.Chung} T.-F. Chung, Y.-L. Yang, B.-M. Huang, Z. Shi, J. Lin, T. Ohmura, J.-R. Yang, {\it Transmission electron microscopy investigation of separated nucleation and in-situ nucleation in {\rm AA7050} aluminium alloy}, \href{https://www.sciencedirect.com/science/article/pii/S1359645418301538}{Acta Materialia 149, 377 (2018)}.

\bibitem{W.Liu} W. Liu, H. Zhao, D. Li, Z. Zhang, G. Huang, Q. Liu, {\it Hot deformation behavior of {\rm AA7085} aluminum alloy during isothermal compression at elevated temperature}, \href{https://www.sciencedirect.com/science/article/pii/S0921509313013750}{Mat. Sc. and Eng. A 596, 176 (2014)}.

\bibitem{B.Sun} B. Sun, Y. Dong, J. Tan, H. Zhang, Y. Sun, {\it Direct observation of $\beta$'' precipitate with a three-dimensional periodic structure in binary $\rm Mg–La$ alloy}, \href{https://www.sciencedirect.com/science/article/pii/S0264127519303776}{Materials and Design 181, 107939 (2019)}.

\bibitem{S.DeWitt} S. DeWitt, E. L.S. Solomon, A. R. Natarajan,
V. Araullo-Peters, S. Rudraraju, L. K. Aagesen, B. Puchala, E. A. Marquis, A. van der Ven, K. Thornton, J. E. Allison, {\it Misfit-driven $\beta$'' precipitate composition and morphology in $\rm Mg–Nd$ alloys}, \href{https://www.sciencedirect.com/science/article/pii/S1359645417305281}{Acta Materialia 136, 378 (2017)}.

\bibitem{E.L.S.Solomon} E. L.S. Solomon, A. R. Natarajan, A. M. Roy, V. Sundararaghavan, A. Van der Ven, E. A. Marquis, {\it Stability and strain-driven evolution of $\beta$' precipitate in $\rm Mg–Y$ alloys}, \href{https://www.sciencedirect.com/science/article/pii/S1359645418309686}{Acta Materialia 166, 148 (2019)}.

\bibitem{P.Ma} P. Ma, C.M. Zou, H.W. Wang, S. Scudino, K.K. Song, M. S. Khoshkhoo,
Z.J. Wei, U. Kuhn, J. Eckert, {\it Structure of GP zones in $\rm AlSi$ matrix composites
solidified under high pressure}, \href{https://www.sciencedirect.com/science/article/pii/S0167577X13008860}{Mater. Lett. 109, 1 (2013)}.

\bibitem{J.Y.Wang} J.-Y. Wang, N. Li, R. Alizadeh, M.A. Monclus, Y.W. Cui, J.M. Molina-Aldareguia, J. LLorca, {\it Effect of solute content and temperature on the deformation mechanisms and critical resolved shear stress in $\rm Mg–Al$ and $\rm Mg–Zn$ alloys}, \href{https://www.sciencedirect.com/science/article/pii/S1359645419301703}{Acta Materialia 170, 155 (2019)}.

\bibitem{D.B.Laks} D.B. Laks, L.G. Ferreira, S. Froyen, A. Zunger, {\it Efficient cluster expansion for substitutional systems}, \href{https://journals.aps.org/prb/abstract/10.1103/PhysRevB.46.12587}{Phys. Rev. B 46, 12587 (1992)}.

\bibitem{M.Asta} M. Asta, C. Wolverton, D. De Fontaine, H Dreysse, {\it Effective cluster interactions from cluster-variation formalism}, \href{https://journals.aps.org/prb/abstract/10.1103/PhysRevB.44.4907}{Phys. Rev. B 44, 4907 (1991)}.

\bibitem{J.H.Chang} J.H. Chang, D. Kleiven, M. Melander, J. Akola, J.M.G. Lastra, T. Vegge, {\it CLEASE: a Versatile and User-friendly Implementation of Cluster Expansion Method}, \href{https://iopscience.iop.org/article/10.1088/1361-648X/ab1bbc/meta}{J. Phys.: Condens. Matter 31, 325901 (2019)}.

\bibitem{J.Zhang} J. Zhang, X. Liu, S. Bi, J. Yin, G. Zhang, M. Eisenbach, {\it Robust data-driven approach for predicting the configurational energy of high entropy alloys}, \href{https://arxiv.org/abs/1908.03665}{arXiv:1908.03665}.

\bibitem{O.I.Gorbatov} O. I. Gorbatov, A.Yu Stroev, Yu. N. Gornostyrev, P. A. Korzhavyi, {\it Effective cluster interactions and preeprecipitate morphology in binary {\rm Al}-based alloys}, \href{https://www.sciencedirect.com/science/article/pii/S135964541930521X}{Acta Materialia 179, 70 (2019)}.




%
%
%
%
%









\bibitem{gp1} J. J. Mortensen, L. B. Hansen, K. W. Jacobsen , {\it Real-space grid implementation of the projector augmented wave method}, \href{https://journals.aps.org/prb/abstract/10.1103/PhysRevB.71.035109}{Phys. Rev. B 71, 035109 (2005)}.

\bibitem{gp2} J. Enkovaara, C. Rostgaard, J.J. Mortensen, J. Chen, M. Dułak, L. Ferrighi, J. Gavnholt, C. Glinsvad, V. Haikola, H.A. Hansen, et al., {\it Electronic structure calculations with GPAW: a real-space implementation of the projector augmented-wave method}, \href{https://iopscience.iop.org/article/10.1088/0953-8984/22/25/253202/meta}{J. Phys. Condens. Matter 22, 253202 (2010)}.

\bibitem{ase} A.H. Larsen, J.J. Mortensen, J. Blomqvist, I.E. Castelli, R. Christensen, M. Dułak, J.Friis, M.N. Groves, B. Hammer, C. Hargus, et al., {\it The atomic simulation environmenta python library for working with atoms}, \href{https://iopscience.iop.org/article/10.1088/1361-648X/aa680e/meta}{J. Phys. Condens. Matter 29, 273002 (2017)}.


\bibitem{N.Metropolis} N. Metropolis and S. Ulam, {\it The Monte Carlo Method}, \href{https://www.tandfonline.com/doi/abs/10.1080/01621459.1949.10483310}{J. Americ. Sta. 44, 247 (1949)}.

\bibitem{S.L.Chen} S.L. Chen, Y.A. Chang, {\it A thermodynamic analysis of the Al-Zn system and phase diagram calculation}, \href{https://www.sciencedirect.com/science/article/pii/036459169390011Y}{CALPHAD 17, 113 (1993)}.

\bibitem{M.Yu.Lavrentiev} M. Yu. Lavrentiev, R. Drautz, D. Nguyen-Manh, T. P. C. Klaver, and S. L. Dudarev, {\it Monte Carlo study of thermodynamic properties and clustering in the bcc Fe-Cr system}, \href{https://journals.aps.org/prb/abstract/10.1103/PhysRevB.75.014208}{Phys. Rev. B 75, 014208 (2007)}.

\bibitem{J.S.Wrobel} J. S. Wrobel, D. Nguyen-Manh, M. Yu. Lavrentiev, M. Muzyk, and S. L. Dudarev, {\it Phase stability of ternary fcc and bcc Fe-Cr-Ni alloys}, \href{https://journals.aps.org/prb/abstract/10.1103/PhysRevB.91.024108}{Phys. Rev. B 91, 024108 (2015)}.















\end{thebibliography}
\end{document}